\documentclass[10pt,journal]{IEEEtran}

\usepackage{cite}
\usepackage[pdftex]{graphicx}
\usepackage{amsmath, amsthm, amssymb, amsfonts}
\usepackage{mathtools}
\usepackage{array}
\usepackage{url}
\usepackage{booktabs}
\usepackage{bm,bbm}
\usepackage{subcaption,subfloat}
\usepackage{orcidlink}
\usepackage{algorithmic}

\newcommand{\short}{RB-SBM}
\newcommand{\full}{Restricted Boltzmann Stochastic Block Model}
\newcommand{\shortmm}{RB-MMSBM}
\newcommand{\fullmm}{Restricted Boltzmann Mixed-Membership Stochastic Block Model}

\newcommand{\bfA}{\mathbf{A}}
\newcommand{\bfB}{\mathbf{B}}
\newcommand{\bfe}{\mathbf{e}}
\newcommand{\bfu}{\mathbf{u}}
\newcommand{\bfv}{\mathbf{v}}
\newcommand{\bfW}{\mathbf{W}}
\newcommand{\bfx}{\mathbf{x}}
\newcommand{\bfy}{\mathbf{y}}
\newcommand{\bfY}{\mathbf{Y}}
\newcommand{\bfz}{\mathbf{z}}
\newcommand{\bfZ}{\mathbf{Z}}
\newcommand{\bfpsi}{\bm{\psi}}
\newcommand{\bfPsi}{\bm{\Psi}}
\newcommand{\bfmu}{\bm{\mu}}

\newcommand{\bbR}{\mathbb{R}}

\newcommand{\rmKL}[2]{\mathrm{KL}(#1 \, \vert\vert \, #2)}
\newcommand{\rmQ}{\mathrm{Q}}
\newcommand{\rmP}{\mathrm{P}}
\newcommand{\rmE}{\mathrm{E}}

\newcommand{\calE}{\mathcal{E}}
\newcommand{\calL}{\mathcal{L}}

\newcommand{\abs}[1]{\vert #1 \vert}

\begin{document}


\title{Equipping SBMs with RBMs: An Explainable Approach for Analysis of Networks with Covariates}

\author{
    Shubham~Gupta~\orcidlink{0000-0001-9692-4096}\,, 
    Gururaj~K~\orcidlink{0000-0001-9235-4815}\,, 
    Ambedkar~Dukkipati~\orcidlink{0000-0002-6352-6283}\,, and 
    Rui~M.~Castro~\orcidlink{0000-0003-4398-0718}

    \IEEEcompsocitemizethanks{
        \IEEEcompsocthanksitem S. Gupta, Gururaj K, and A. Dukkipati are with the Department of Computer Science and Automation, Indian Institute of Science, Bengaluru, Karnataka, India, 560012.\protect~E-mail: \{shubhamg, gururajk, ambedkar\}@iisc.ac.in
        \IEEEcompsocthanksitem Rui M. Castro is with Department of Mathematics and Computer Science, Eindhoven University of Technology, Eindhoven  5600  MB,  The  Netherlands.\protect~E-mail: r.m.pires.da.silva.castro@tue.nl
    }

    \thanks{}
}

\markboth{}{Gupta \MakeLowercase{\text
it{et al.}}: Equipping SBMs with RBMs: An Explainable Approach for Analysis of Networks with Covariates}


\IEEEtitleabstractindextext{%
\begin{abstract}
    Networks with node covariates offer two advantages to community detection methods, namely, (i) exploit covariates to improve the quality of communities, and more importantly, (ii) explain the discovered communities by identifying the relative importance of different covariates in them. Recent methods have almost exclusively focused on the first point above. However, the quantitative improvements offered by them are often due to complex black-box models like deep neural networks at the expense of explainability. Approaches that focus on the second point are either domain-specific or have poor performance in practice. This paper proposes explainable, domain-independent statistical models for networks with node covariates that additionally offer good quantitative performance. Our models combine the strengths of Stochastic Block Models and Restricted Boltzmann Machines to provide interpretable insights about the communities. They support both pure and mixed community memberships. Besides providing explainability, our approach's main strength is that it does not explicitly assume a causal direction between community memberships and node covariates, making it applicable in diverse domains. We derive efficient inference procedures for our models, which can, in some cases, run in linear time in the number of nodes and edges. Experiments on several synthetic and real-world networks demonstrate that our models achieve close to state-of-the-art performance on community detection and link prediction tasks while also providing explanations for the discovered communities.
\end{abstract}

\begin{IEEEkeywords}
Explainability, Interpretability, Covariates, Stochastic Block Models, Community Detection, Link Prediction
\end{IEEEkeywords}}

\maketitle
\IEEEdisplaynontitleabstractindextext

    
\section{Introduction}
\label{sec:introduction}
\IEEEPARstart{M}{any} real-world networks exhibit a community structure. Communities summarize networks by identifying similar nodes in them (e.g., grouping webpages by topic on the internet) \cite{Fortunato:2010:CommunityDetectionInGraphs}. Finding communities is useful in many applications like recommendation systems \cite{LindenEtAl:2003:AmazonRecommendations,WuEtAl:2015:ClusteringAndInferenceFromPairwiseComparisons}, image segmentation \cite{SnijdersEtAl:1997:EstimationAndPredictionForSBMsForGraphsWithLatentBlockStructure,GhoshdastidarEtAl:2015:AProvableGeneralizedTensorSpectralMethodForUniformHypergraphPartitioning}, and market research \cite{ClausetEtAl:2004:FindingCommunityStructureInVeryLargeNetworks}, to name a few. 

The formalization of a community detection problem requires several concrete choices, such as how similarities between nodes are quantified \cite{Fortunato:2010:CommunityDetectionInGraphs}. Thus, methods for finding communities should explain their output by identifying the properties of the members based on which they were grouped. Traditionally, researchers have used metrics like the modularity score \cite{Newman:2018:Networks} to judge the quality of the discovered communities. These metrics usually have a notion of an ideal community, and they score the algorithm's output against this notion. However, they do not provide any insights about the algorithm's decision to group nodes the way it has done. Networks with node covariates present an opportunity to address this issue.

A prototypical network consists of nodes that encode subjects (e.g., people, webpages) and links between them (e.g., a friendship relation or a link between two webpages). More often than not, there is also extra information associated with each node or link (e.g., gender and age of a person or the topic of a webpage). In the literature, these networks are known as \textit{attributed networks} or \textit{networks with covariates} \cite{YangEtAl:2013:CommunityDetectionInNetworksWithNodeAttributes,BinkiewiczEtAl:2017:CovariateAssistedSpectralClustering}. Traditionally, node covariates have been used by community detection methods to improve their performance, especially on sparse networks where the connectivity pattern encodes limited information. However, a method can also use these covariates to explain its output by identifying the defining covariates of the discovered communities\footnote{\textit{Defining covariates} refer to the covariates that the method has used for justifying its decision to group the nodes together. This is different from identifying the shared covariates in the communities after their discovery. The covariates identified post community discovery may just be correlated with the communities without being causally responsible for the method's output.}. For example, in a social network, a method can explain its decision to group certain people together by highlighting that they have similar ages and have attended the same high school.

Existing approaches for attributed networks can be divided into two categories. In the first category, methods focus on using covariates to improve the quality of communities \cite{RuanFuhryEtAl:2013:EfficientCommunityDetectionInLargeNetworksUsingContentAndLinks,PerozziEtAl:2014:DeepWalk,KipfEtAl:2016:VariationalGraphAutoEncoders,PanEtAl:2018:AdversariallyRegularizedGraphAutoencoderForGraphEmbedding,WangEtAl:2019:AttributedGraphClustering,MehtaEtAl:2019:SBMsMeetGNN}. Many recent approaches belong to this category and use complex neural networks for quantitative gains in performance \cite{KipfEtAl:2016:VariationalGraphAutoEncoders,PanEtAl:2018:AdversariallyRegularizedGraphAutoencoderForGraphEmbedding,WangEtAl:2019:AttributedGraphClustering,MehtaEtAl:2019:SBMsMeetGNN}. However, the outputs of these black-box methods are often hard to interpret. As argued above, interpreting or explaining the communities is often as crucial in practice as quantitative performance. The second category includes statistical models for attributed networks \cite{CohnHofmann:2001:TheMissingLinkAProbabilisticModelOfDocumentContentAndHypertextConnectivity,EroshevaEtAl:2004:MixedMembershipModelsOfScientificPublications, LiuEtAl:2009:TopicLinkLDA, ChangBlei:2009:RelationalTopicModelingForDocumentNetworks, BalasubramanyanEtAl:2011:BlockLDA, YangEtAl:2013:CommunityDetectionInNetworksWithNodeAttributes}. While these models often have the advantage of explainability, many of them suffer from poor practical performance (see Section \ref{section:experiments}). Moreover, they are often designed for a particular domain (such as document networks) and have limited applicability.

In contrast, we develop explainable and domain-independent statistical models for attributed networks. Experiments demonstrate that our models often perform on par with (and in some cases even better than) more complex neural-network-based methods while being more explainable. Next, we briefly introduce the proposed models and highlight our contributions in making them explainable and domain-independent.

We propose two models that combine variants of the stochastic block model (SBM) \cite{HollandEtAl:1983:StochasticBlockmodelsFirstSteps} and restricted Boltzmann machines (RBM) \cite{FischerEtAl:2012:AnIntroductionToRestrictedBoltzmannMachines}. An SBM samples the community membership of nodes from a multinomial distribution. Our first model integrates node covariates by using a RBM to model the distribution of community membership given the node covariates, which is then used together with a SBM model. We call this model \full{} (\short{}). Our next model extends this idea to support mixed community memberships where a node can simultaneously belong to multiple communities. It uses a modified variant of RBMs with a Mixed-Membership SBM \cite{Airoldi:2008:MixedMembershipStochasticBlockmodels} in a similar manner as above. We call this \fullmm{} (\shortmm{}). 

We subscribe to the view that one can explain a community by specifying its relationship with other communities and identifying salient covariates that its members share. An SBM (or its variant) specifies how the communities are related to each other. Similarly, an RBM (or its variants) offers insights by identifying the node covariates that play a vital role in each community. These two components are relatively simple and easy to interpret, which is helpful to get qualitative insights about the discovered communities. For example, in sequel, we demonstrate that our method can identify meaningful communities in a network of web pages pertaining to philosophers on Wikipedia, while also highlighting the covariates (words on the webpages) that were responsible for the method's decision to group the philosophers together (see Figure~\ref{fig:philosophers} for details). These covariate-based explanations for the model's decisions are absent in modern deep learning-based methods. Moreover, combining these two seemingly simple modeling components still leads to nearly state-of-the-art performance in many realistic settings.

To understand why most existing statistical models are domain-dependent, note that they usually factorize the distribution over node covariates and community memberships into two terms. The first term assigns a probability to the edges using nodes' community memberships (e.g., SBM). The second term captures the relation between node covariates and community membership (e.g., LDA). Existing approaches often model the second term by using a conditional distribution between attributes and communities \cite{YangEtAl:2013:CommunityDetectionInNetworksWithNodeAttributes,EroshevaEtAl:2004:MixedMembershipModelsOfScientificPublications,ChangBlei:2009:RelationalTopicModelingForDocumentNetworks,MehtaEtAl:2019:SBMsMeetGNN}. The ease of modeling in an application domain dictates the choice of the conditioning variable (explicitly modeling covariates conditioned on the community membership or vice versa), typically making the proposed models rather domain-specific. In general, this choice of ``direction'' in the conditional distribution can be challenging to make. We avoid making such a decision explicitly by modeling a joint distribution between node covariates and communities using an RBM. As RBMs can represent a rich class of joint distributions, our models have wide applicability. However, one can always interpret the RBM as modeling a conditional distribution, if desired.

We also derive efficient inference procedures for our models using the variational Expectation-Maximization strategy. Notably, for RB-SBM, each iteration in the inference procedure runs in linear time in the number of nodes and edges. The complexity of inference in RB-MMSBM matches that of the mixed-membership SBM. Note that both RB-SBM and RB-MMSBM are generative. Hence, they can also be used to sample random attributed networks with various properties, which can be useful for testing the performance of other community detection algorithms. Finally, while our discussion until this point has focused on community detection, experiments show that our models also offer good link-prediction performance.

\textbf{Contributions and organization}: \textbf{(i)} We propose two domain-independent and easily interpretable statistical models for networks with covariates, respectively \short{} (see Section \ref{section:rbsbm}) and \shortmm{} (see Section \ref{section:rbmmsbm}). \textbf{(ii)} We develop practical and efficient inference methods for the proposed models (Sections \ref{section:inference_in_rbsbm} and \ref{section:inference_in_rbmmsbm}). In particular, each iteration of the inference algorithm for \short{} runs in linear time in the number of nodes and edges, thus making it scalable to large networks. 
\textbf{(iii)} We empirically validate the proposed models on community detection and link prediction tasks (Section \ref{section:experiments}). Our models set benchmarks for performance on Cora and Citeseer networks in terms of NMI score with respect to known ground-truth communities. They also yield qualitative insights about the communities by identifying the key covariates in them.


\section{Related Work}
\label{section:related_work}

Approaches for community detection in networks with covariates can be broadly divided into three categories: \textbf{(i)} those that modify algorithms for networks without covariates to include covariates, \textbf{(ii)} statistical models, and \textbf{(iii)} deep neural network-based approaches. This paper belongs to the second category.

Notable examples from the first category include \cite{ZhouEtAl:2010:ClusteringLargeAttributedGraphs,RuanFuhryEtAl:2013:EfficientCommunityDetectionInLargeNetworksUsingContentAndLinks}, and \cite{BinkiewiczEtAl:2017:CovariateAssistedSpectralClustering}. In \cite{ZhouEtAl:2010:ClusteringLargeAttributedGraphs,RuanFuhryEtAl:2013:EfficientCommunityDetectionInLargeNetworksUsingContentAndLinks}, the authors use node covariates to augment the set of links in the network. Existing community detection algorithms are then used on this augmented network. \cite{BinkiewiczEtAl:2017:CovariateAssistedSpectralClustering} proposed Covariate Assisted Spectral Clustering (CASC) that uses spectral clustering algorithm \cite{Luxburg:2007:ATutorialOnSpectralClustering} on a similarity matrix that combines both node covariates and connectivity information. Spectral clustering has been adapted to networks with covariates in other ways as well, see references within \cite{BinkiewiczEtAl:2017:CovariateAssistedSpectralClustering} for more details. Other approaches modify the objective function of an existing algorithm to include node covariates. For example, \cite{ZhangEtAl:2016:CommunityDetectionInNetworksWithNodeFeatures} propose \textit{joint community detection criteria}, which modifies the well-known modularity score to include covariates. See references within \cite{LiEtAl:2011:GeneralizedLatentFactorModelsForSocialNetworkAnalysis,AkogluEtAl:2012:PICS,ZhangEtAl:2016:CommunityDetectionInNetworksWithNodeFeatures} for other similar approaches. Along similar lines, \cite{LiEtAl:2018:CommunityDetectionInAttributedGraphsAnEmbeddingsApproach} proposed \textit{Community Detection in attributed graphs: an Embedding approach} (CDE) that relies on simultaneous non-negative matrix factorization of node-covariate and adjacency matrices. Often these modifications are done in \textit{ad-hoc} ways, making it hard to reason about the discovered communities.

In general, approaches based on statistical models offer more insights as compared to purely algorithmic approaches. For example, using our approach, one can naturally find the role played by different node covariates in characterizing various communities. Both discriminative \cite{YangEtAl:2009:CombiningLinkAndContentForCommunityDetection} and generative probabilistic models \cite{CohnHofmann:2001:TheMissingLinkAProbabilisticModelOfDocumentContentAndHypertextConnectivity,EroshevaEtAl:2004:MixedMembershipModelsOfScientificPublications,ChangBlei:2009:RelationalTopicModelingForDocumentNetworks,LiuEtAl:2009:TopicLinkLDA,BalasubramanyanEtAl:2011:BlockLDA,XuEtAl:2012:AModelBasedApproachToAttributedGraphClustering,YangEtAl:2013:CommunityDetectionInNetworksWithNodeAttributes} have been proposed. However, most statistical models tend to be domain-specific and are only applicable to a small class of networks, like document networks. Methods for document networks usually employ variants of Latent Dirichlet Allocation (LDA) \cite{BleiNgJordan:2003:LatentDirichletAllocation} to model textual node covariates. For example, \cite{EroshevaEtAl:2004:MixedMembershipModelsOfScientificPublications} proposed a model that we will call LDA-Link-Word (LLW) following \cite{YangEtAl:2009:CombiningLinkAndContentForCommunityDetection}. This model uses LDA for modeling documents and uses the notion of communities for modeling links. In \cite{YangEtAl:2009:CombiningLinkAndContentForCommunityDetection} the authors use a node popularity based conditional link model (PCL) and combine it with PLSA (Probabilistic Latent Semantic Analysis, which is similar to LDA) to model documents. They call this model PCL-PLSA. They also have a discriminative variant of the model which replaces PLSA with a discriminative content model (PCL-DC). A notable exception to being domain specific is Community Detection in Networks with Node covariates (CESNA) \cite{YangEtAl:2013:CommunityDetectionInNetworksWithNodeAttributes} which supports binary node covariates and overlapping communities. We demonstrate our approach outperforms all of these approaches while retaining their explainability.

More recently, several deep neural network-based approaches have been proposed, \cite{KipfEtAl:2016:VariationalGraphAutoEncoders,PerozziEtAl:2014:DeepWalk, PanEtAl:2018:AdversariallyRegularizedGraphAutoencoderForGraphEmbedding, WangEtAl:2019:AttributedGraphClustering, MehtaEtAl:2019:SBMsMeetGNN, DiEtAl:2019:CommunityDetectionViaJointGCNEmbeddingInAttributeNetwork}. Some of these works include but are not limited to generative models like \cite{MehtaEtAl:2019:SBMsMeetGNN} and discriminative models like \cite{HamiltonEtAl:2017:InductiveRepresentationLearningOnLargeGraphs, DiEtAl:2019:CommunityDetectionViaJointGCNEmbeddingInAttributeNetwork, VelicovicEtAl:2019:DeepGraphInfomax}. One of the main motivation for our work is to come up with a simple generative model for networks with covariates that supports a lightweight inference procedure, while offering explainable insights about the discovered communities. Compared to neural network based methods, our models: \textbf{(i)} do not require complex neural network architectures to achieve benchmark performance, \textbf{(ii)} support a tractable inference procedure with fewer trainable parameters, \textbf{(iii)} yield explainable insights about communities and the role of node covariates in defining these communities. In contrast, the above-mentioned approaches are more complex and usually lack interpretability.

All the approaches above (except \cite{YangEtAl:2013:CommunityDetectionInNetworksWithNodeAttributes}) assume that nodes belong to exactly one community. An extension of the stochastic block model, called Mixed Membership SBM \cite{Airoldi:2008:MixedMembershipStochasticBlockmodels}, relaxes this assumption. Mixed-membership SBM has been adapted in various ways by considering different priors like copula functions \cite{FanEtAl:2016:CopulaMixedMembershipSBM} and logistic functions \cite{KimLeskovec:2012:LatentMultiGroupMembershipGraphModel}. In contrast, our model \shortmm{} incorporates node covariates in a mixed-membership SBM. As before, it retains the explainability of mixed-membership SBM, while offering good practical performance. We have restricted our attention to SBM and its variants as SBM is an expressive model that has been extensively studied \cite{RoheEtAl:2011:SpectralClusteringAndTheHighDimensionalSBM,LeiEtAl:2015:ConsistencyOfSpectralClusteringInSBM} in several different settings \cite{KarrerNewman:2011:StochasticBlockmodelsAndCommunityStructureInNetworks, GhoshdastidarDukkipati:2017:ConsistencyOfSpectralHypergraphPartitioningUnderPlantedPartitionModel}. This allows the techniques that we have developed to be extended to other variants of SBM, providing interesting directions for future work.


\section{\short}
\label{section:rbsbm}

\paragraph*{Notation} We consider (directed or undirected) simple networks with $n$ nodes. These are described by an adjacency matrix $\bfA \in \{0, 1\}^{n \times n}$, where $A_{ii} = 0$ for all $i = 1, \dots, n$. We additionally assume that each node has a vector of $m$ covariates associated with it. For the $i^{th}$ node, this vector is denoted by $\bfy_i \in \bbR^m$. Let $\bfY \in \bbR^{n \times m}$ be the matrix with $\bfy_1, \dots, \bfy_n$ as its rows. Our models assume a latent community structure in the network with $k$ communities. Nodes have a latent community membership vector $\bfz_i \in \bbR^k$, which is one-hot encoded for pure community memberships ($z_{ij} = 1$ if $i^{th}$ node belongs to $j^{th}$ community and $z_{ij} = 0$ otherwise) and $\bfz_i \in \Delta_k$ for mixed-community membership. Here, $\Delta_k \coloneqq \{\bfx \in \bbR^k : \forall i, x_i \geq 0 \text{ and } \sum_{i=1}^k x_i = 1\}$ is the $k$-dimensional simplex. Analogously as above define $\bfZ \in \bbR^{n \times k}$ to be a matrix with $\bfz_1, \dots, \bfz_n$ as its rows.


This section describes \short{} for modeling networks with binary node covariates where each node belongs to exactly one community. We describe the model in Section \ref{section:rbsbm_description} and develop an efficient inference procedure in Section \ref{section:inference_in_rbsbm}. The use of binary covariates simplifies some aspects of the model but is not a severe restriction. This assumption is lifted in Section~\ref{section:continuous_rbsbm}.


\subsection{Model Description}
\label{section:rbsbm_description}

\begin{figure}
    \includegraphics[width=\linewidth]{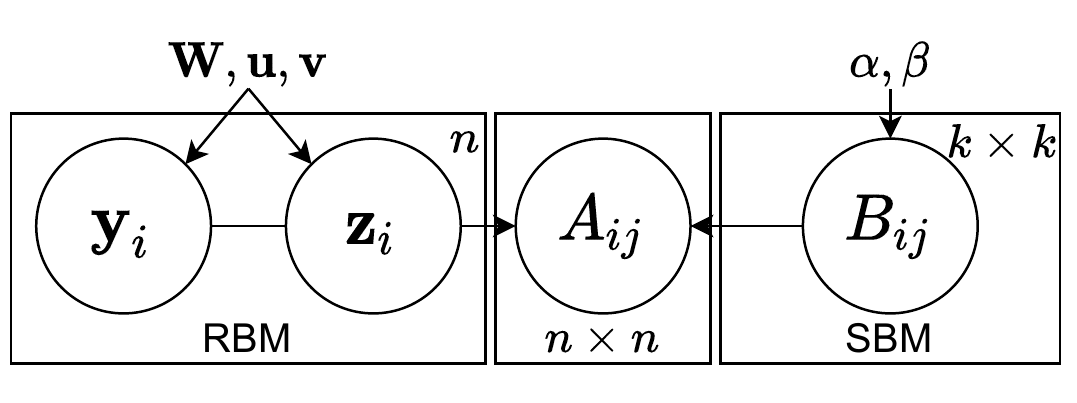}
    \caption{Graphical model for \short{}}
    \label{fig:graphical_model_rbsbm}
\end{figure}

\short{} uses an RBM to model a joint distribution between covariates $\bfY$ and communities $\bfZ$. It then combines it with an SBM to specify the probability of an observed network $\bfA$. These two components are described in detail below.

First, we modify an RBM to encode a joint distribution between binary covariate vectors $\bfy \in \{0, 1\}^m$ and $k$ dimensional one-hot encoded community membership vectors $\bfz$ as
\begin{equation}
    \label{eq:prob_y_z}
    \rmP_{\bm{\theta}}(\bfy, \bfz) = \frac{\exp(\bfy^\intercal \bfW \bfz + \bfy^\intercal \bfu + \bfz^\intercal \bfv)}{\Psi(\bfW, \bfu, \bfv)}.
\end{equation}
Here, $\bm{\theta} = \{\bfW, \bfu, \bfv\}$ are the parameters of the RBM and $\Psi(\bfW, \bfu, \bfv)$ is the normalization constant given by
\begin{equation}
    \label{eq:partition_function_computation_main}
    \Psi(\bfW, \bfu, \bfv) = \sum_{i=1}^k \exp{(v_i)} \prod_{j=1}^m (1 + \exp(W_{ji} + u_j)).
\end{equation}
The derivation of \eqref{eq:partition_function_computation_main} is given in Appendix \ref{appendix:normalization_constant_rbm}. We further assume that $(\bfy_i, \bfz_i)$ pairs are drawn independently for all nodes $i = 1, \dots, n$, hence $\rmP_{\bm{\theta}}(\bfY, \bfZ) = \prod_{i = 1}^n \rmP_{\bm{\theta}}(\bfy_i, \bfz_i)$. Simple calculations show that the conditional distributions are given by
\begin{equation}
\label{eq:conditional_distributions}
\begin{aligned}
    \rmP_{\bm{\theta}}(y_j = 1 \,\vert\, \bfz) &= \frac{1}{1 + \exp(-\sum_{\ell = 1}^k z_{\ell} W_{j\ell} - u_j)}, \\
    \rmP_{\bm{\theta}}(z_\ell = 1 \,\vert\, \bfy) &= \frac{\exp(\sum_{j = 1}^m y_j W_{j\ell} + v_\ell)}{\sum_{\ell^{'} = 1}^k \exp(\sum_{j = 1}^m y_j W_{j\ell^{'}} + v_{\ell^{'}})}.
\end{aligned}
\end{equation}
These conditional distributions can be used to draw samples from $\rmP_{\bm{\theta}}(\bfy, \bfz)$ via Gibbs sampling.

Second, we use an SBM to model the edges in the network. Let $\bfB \in [0,1]^{k \times k}$ denote a block matrix specifying the probability with which nodes in various communities connect. In \short{}, conditioned on the community membership of nodes $\bfZ$, each edge in $\bfA$ is modeled as an independent Bernoulli random variable. In particular, the probability of an edge between nodes $i$ and $j$ is given by $\rmP(A_{ij} = 1 \vert \bfz_i, \bfz_j, \bfB) = \bfz_i^\intercal \bfB \bfz_j$. Often in a traditional SBM model the community membership $\bfz_i$ of nodes are sampled from a multinomial distribution. However, in \short{}, we sample $(\bfy_i, \bfz_i)$ pairs from the RBM as mentioned above. We take a Bayesian perspective and impose a \textrm{Beta} prior on all entries in $\bfB$. Namely, each entry $B_{ij}$ is an independent sample from a \textrm{Beta} distribution with parameters $\alpha_{ij}, \beta_{ij} > 0$, that are specified by the user.

A high-level graphical description of the model is provided in Figure~\ref{fig:graphical_model_rbsbm}. The generative process of \short{} is summarized below.
\begin{enumerate}
    \item Sample $B_{ij} \sim \textrm{Beta}(\alpha_{ij}, \beta_{ij})$ for $i, j = 1, \ldots, k$
    \item Sample $(\bfy_i, \bfz_i)$ for $i = 1, \dots, n$ from the RBM
    \item Sample $A_{ij} \sim \textrm{Bernoulli}(\bfz_i^\intercal \bfB \bfz_j)$ for all $i \neq j$ from the SBM
\end{enumerate}

Using the independence assumptions implied by Figure~\ref{fig:graphical_model_rbsbm}, the probability density function\footnote{Formally this is the Radon-Nikodym derivative with respect to a dominating measure consisting of the product of counting measures (for the first three arguments) and Lebesgue measure (for the last argument).} can be factorized as:
\begin{equation}
    \label{eq:joint_distr_compressed}
    \rmP_{\bm{\theta}}(\bfA, \bfY, \bfZ, \bfB) = \rmP_{\bm{\theta}}(\bfB) \, \rmP_{\bm{\theta}}(\bfY, \bfZ) \, \rmP_{\bm{\theta}}(\bfA \vert \bfB, \bfZ).
\end{equation}
Consequently, the log-probability, $L_{\bm{\theta}} \coloneqq \ln \rmP_{\bm{\theta}}(\bfA, \bfY, \bfZ, \bfB)$, becomes:
{
\small
\begin{equation}
\label{eq:log_joint}
\begin{aligned}
	L_{\bm{\theta}} &= \ln \rmP_{\bm{\theta}}(\bfB) + \ln \rmP_{\bm{\theta}}(\bfY, \bfZ) + \ln \rmP_{\bm{\theta}}(\bfA \vert \bfB, \bfZ) \\
	&= \sum_{i, j = 1}^k (\alpha_{ij} - 1) \ln B_{ij} +  (\beta_{ij} - 1) \ln (1 - B_{ij}) \\
	&\;\;\; - \sum_{i, j = 1}^k \ln B(\alpha_{ij}, \beta_{ij}) - n \ln \Psi(\bfW, \bfu, \bfv) \\
	&\;\;\; + \sum_{i=1}^{n} \Big(\sum_{j=1}^m \sum_{\ell=1}^k Y_{ij} Z_{i\ell} W_{j\ell} + \sum_{j=1}^m u_j Y_{ij} + \sum_{\ell=1}^k v_\ell Z_{i\ell}\Big) \\
	&\;\;\; + \sum_{i \neq j}\sum_{\ell_1, \ell_2 = 1}^k \Big( A_{ij} Z_{i\ell_1} Z_{j\ell_2} \ln B_{\ell_1 \ell_2} + \\
    &\hspace{2.3cm} (1 - A_{ij}) Z_{i\ell_1} Z_{i\ell_2} \ln (1 -B_{\ell_1 \ell_2}) \Big).
\end{aligned}
\end{equation}
}%
Here, $B(.)$ is the Beta function. In addition to $\bm{\theta} = \{\bfW, \bfu, \bfv\}$, the model uses three hyperparameters - $\bm{\alpha}$, $\bm{\beta}$ and $k$ that are provided by the user. The first two describe the prior on $\bfB$ and $k$ as the number of communities. Although $L_{\bm{\theta}}$ depends on the hyperparameters as well, we have suppressed this in the notation to avoid unnecessary clutter. Practical ways to set hyperparameters are discussed in Section \ref{section:experiments}. A few remarks about \short{} are in order:

\paragraph{Model explainability} As noted before, \short{} can explain the communities inferred by it. The entries of matrix $\bfW$ specify the relation between various covariates and communities. Namely, a high value of $W_{j\ell}$ indicates that a node $i$ for which $Y_{ij} = 1$ has an increased likelihood of belonging to the $\ell^{th}$ community, thus associating the $j^{th}$ covariate to the $\ell^{th}$ community. Similarly, the entries of $\bfB$ specify the relationship between communities.

\paragraph{Modeling joint distribution $\rmP_{\bm{\theta}}(\bfy, \bfz)$} In contrast with many approaches in the literature (see Section \ref{section:related_work}), \short{} does not explicitly regard that the covariates are dictated by or dictate the community structure. While such assumptions can be reasonable in certain settings, they do not always hold in practice. For example, in a social network, while people become part of communities based on shared attributes (modeling $\rmP(\bfz \vert \bfy)$), they also acquire attributes like preferences for music genres due to their community membership (modeling $\rmP(\bfy \vert \bfz)$). Modeling the joint distribution naturally captures both types of interactions between $\bfy$ and $\bfz$ and makes \short{} compatible with several application domains.

\paragraph{Representation power} Both RBM and SBM have good representation power and hence our model can be used to generate a large class of networks with various properties. For example, setting the diagonal entries of $\bfB$ higher than the other entries leads to traditional assortative communities. One can similarly impose other structures on $\bfB$ to have hierarchical communities, disassortative communities, etc.

\paragraph{Computation of the normalization constant} In general, the computation of normalization constant $\Psi(\bfW, \bfu, \bfv)$ in RBMs is intractable. However, this computation is greatly simplified in our case because $\bfz_i$ are one-hot encoded vectors, and one can efficiently compute $\Psi(\bfW, \bfu, \bfv)$ in $O(nm)$ operations (see Appendix \ref{appendix:normalization_constant_rbm}). This has a significant impact in the computational complexity of the inference algorithms.

\paragraph{Reduction to SBM} \short{} reduces to an SBM with parameters $(\bm{\pi}, \bfB)$ by setting $\bfW = \mathbf{0}$ and $v_i = \ln \pi_i$ for $i = 1, \dots, k$.

Despite its flexibility, \short{} retains a lightweight inference procedure described in the next section.


\subsection{Inference in \short{}}
\label{section:inference_in_rbsbm}

Given a network with covariates, the only observations are the connectivity structure $\bfA$ and node covariates $\bfY$, while the community membership $\bfZ$ and block structure $\bfB$ are hidden from us. Inference is concerned with computing $\rmP_{\bm{\theta}}(\bfZ, \bfB \vert \bfA, \bfY)$ and estimating parameters $\bm{\theta}$ to perform tasks like link prediction and community detection. As calculating $\rmP_{\bm{\theta}}(\bfZ, \bfB \vert \bfA, \bfY)$ requires a summation over $k^n$ choices of community assignments, performing exact inference in \short{} is intractable. Therefore, we resort to approximate inference techniques and use variational inference \cite{BleiEtAl:2017:VariationalInferenceAReviewForStatisticians}. More specifically, we use a variational EM algorithm that alternates between the following two steps \cite{Bishop:2006:PRML}:
\begin{enumerate}
    \item \textbf{E-step:} Find an approximate posterior distribution over $\bfZ$ and $\bfB$ assuming knowledge of $\bm{\theta}$.
    \item \textbf{M-step:} Find a point estimate of $\bm{\theta}$ assuming the distribution over $\bfZ$ and $\bfB$ is the one obtained in step 1.
\end{enumerate}
Next, we describe these two steps.

Let $\calL(\bm{\theta}) = \ln \rmP_{\bm{\theta}}(\bfA, \bfY)$ be the log-probability of the observed data. We maximize a lower bound on $\calL(\bm{\theta})$ instead of maximizing $\calL(\bm{\theta})$ itself, as its computation is intractable. Let $\rmQ(\bfZ, \bfB)$ be a distribution that approximates the true posterior $\rmP_{\bm{\theta}}(\bfZ, \bfB \vert \bfA, \bfY)$. Then,
\begin{align*}
    \label{eq:lower_bound_derivation}
    \calL(\bm{\theta}) &= \rmE_{\rmQ}[\ln \rmP_{\bm{\theta}}(\bfA, \bfY)] \\
    &= \rmE_{\rmQ}[\ln \rmP_{\bm{\theta}}(\bfA, \bfY, \bfZ, \bfB) - \ln \rmP_{\bm{\theta}}(\bfZ, \bfB \vert \bfA, \bfY)] \\
    &= \rmE_{\rmQ}[\ln \frac{\rmP_{\bm{\theta}}(\bfA, \bfY, \bfZ, \bfB)}{\rmQ(\bfZ, \bfB)} - \ln \frac{\rmP_{\bm{\theta}}(\bfZ, \bfB \vert \bfA, \bfY)}{\rmQ(\bfZ, \bfB)}] \\
    &= \calL_{\rmQ}(\bm{\theta}) + \rmKL{\rmQ(\bfZ, \bfB)}{\rmP_{\bm{\theta}}(\bfZ, \bfB | \bfA, \bfY))} \\
    &\geq \calL_{\rmQ}(\bm{\theta}),
\end{align*}
where, $\calL_{\rmQ}(\bm{\theta}) = \rmE_{\rmQ}[\ln \rmP_{\bm{\theta}}(\bfA, \bfY, \bfZ, \bfB) - \ln \rmQ(\bfZ, \bfB)]$ and $\rmKL{\cdot}{\cdot}$ is the Kullback-Leibler divergence. The bound $\calL_{\rmQ}(\bm{\theta})$ is generally denoted as the Evidence Lower Bound (\textrm{ELBO}) \cite{BleiEtAl:2017:VariationalInferenceAReviewForStatisticians}. In the E-step, $\bm{\theta}$ is held constant and the \textrm{ELBO} is maximized over $\rmQ$, and in the M-step, $\rmQ$ is held constant and \textrm{ELBO} is maximized over $\bm{\theta}$.


\subsubsection{E-step}
\label{section:e_step}

As is common in practice, we assume that $\rmQ$ belongs to the mean-field family of distributions \cite{BleiEtAl:2017:VariationalInferenceAReviewForStatisticians}, i.e., 
\begin{equation}
    \label{eq:mean_field}
    \rmQ(\bfZ, \bfB) = \Big(\prod_{i=1}^n q_i(\bfz_i)\Big) \Big(\prod_{i, j = 1}^k q_{ij}(B_{ij})\Big).
\end{equation}
where $q_{ij}(\cdot)$ and $q_{i}(\cdot)$ are arbitrary distributions over which the optimization is performed.
Using coordinate ascent to approximately maximize $\calL_{\rmQ}(\bm{\theta})$ with respect to $\rmQ$ yields distributions $q^*_{ij}$ and $q^*_i$ that have the following form (see also \cite{BleiEtAl:2017:VariationalInferenceAReviewForStatisticians})
\begin{equation}
\label{eq:cavi_updates}
\begin{aligned}
    q^*_{ij}(B_{ij}) &\propto \exp(\rmE_{\rmQ_{-ij}}[\ln \rmP_{\bm{\theta}}(\bfA, \bfY, \bfZ, \bfB)]) \text{ and} \\
    q^*_i(\bfz_i) &\propto \exp(\rmE_{\rmQ_{-i}}[\ln \rmP_{\bm{\theta}}(\bfA, \bfY, \bfZ, \bfB)]),
\end{aligned}
\end{equation}
where $\rmE_{\rmQ_{-ij}}[.]$ and $\rmE_{\rmQ_{-i}}[.]$  represent the expectation with respect to all distributions on the right hand side of \eqref{eq:mean_field} except $q_{ij}$ and $q_{i}$, respectively. We often use $q_{i}(j)$ as a shorthand for $q_i(Z_{ij} = 1)$. Next, we compute the expectations in \eqref{eq:cavi_updates}.
{
\small
\begin{equation}
\label{eq:exp_log_joint_qkl}
\begin{aligned}
	\rmE&_{\rmQ_{-ij}}[\ln \rmP_{\bm{\theta}}(\bfA, \bfY, \bfZ, \bfB)] = (\alpha_{ij} - 1)\ln B_{ij} \\
    &+  (\beta_{ij} - 1)\ln (1 - B_{ij}) + \sum_{i_1 \neq i_2} \Big[ A_{i_1i_2} q_{i_1}(i) q_{i_2}(j) \ln B_{ij} \\
	&+ (1 - A_{i_1i_2}) q_{i_1}(i) q_{i_2}(j) \ln(1 - B_{ij}) \Big] + \textrm{const}_1.
\end{aligned}
\end{equation}
}%
The literal $\textrm{const}_1$ contains all terms that do not depend on $B_{ij}$. Using \eqref{eq:exp_log_joint_qkl} in \eqref{eq:cavi_updates}, we get
{
\small
\begin{align*}
	q^*_{ij}&(B_{ij}) \propto \exp\Big[\Big(\alpha_{ij} - 1 + \sum_{i_1 \neq i_2} A_{i_1i_2} q_{i_1}(i) q_{i_2}(j)\Big) \ln B_{ij} \\
	&\;\;\;+ \Big(\beta_{ij} - 1 + \sum_{i_1 \neq i_2} (1 - A_{i_1i_2}) q_{i_1}(i) q_{i_2}(j)\Big) \ln(1 -B_{ij}) \Big].
\end{align*}
}%
Notice that this is the exponential family form of a \textrm{Beta} distribution. Thus, $q^*_{ij}$ is $\textrm{Beta}(\bar{\alpha}_{ij}, \bar{\beta}_{ij})$, where
\begin{align*}
    \bar{\alpha}_{ij} &= {\alpha}_{ij} + \sum_{i_1 \neq i_2} A_{i_1 i_2} q_{i_1}(i) q_{i_2}(j) \text{ and}\\
    \bar{\beta}_{ij} &= {\beta}_{ij} + \sum_{i_1 \neq i_2} (1 - A_{i_1i_2}) q_{i_1}(i) q_{i_2}(j).
\end{align*}

Similarly, for the second expectation in \eqref{eq:cavi_updates}, let $\bm{\Psi}(\cdot)$ denote the digamma function. Using \eqref{eq:log_joint}, \eqref{eq:mean_field}, and linearity of expectations, we get
{
\small
\begin{equation}
    \label{eq:exp_log_joint_qn}
\begin{aligned}
	\rmE&_{\rmQ_{-i}}[\ln \rmP_{\bm{\theta}}(\bfA, \bfY, \bfZ, \bfB)] = \sum_{j=1}^m \sum_{\ell=1}^k Y_{ij} Z_{i\ell} W_{j\ell} + \sum_{\ell=1}^k v_\ell Z_{i\ell} \\
	& + \sum_{j \neq i}\sum_{\ell_1, \ell_2 = 1}^k q_{j\ell_1} A_{ji} Z_{i\ell_2} (\bm{\Psi}(\bar{\alpha}_{\ell_1\ell_2}) - \bm{\Psi}(\bar{\alpha}_{\ell_1\ell_2} + \bar{\beta}_{\ell_1\ell_2})) \\
	& + \sum_{j \neq i} \sum_{\ell_1, \ell_2 = 1}^k q_{j\ell_1} (1 - A_{ji})  Z_{i\ell_2} (\bm{\Psi}(\bar{\beta}_{\ell_1\ell_2}) - \bm{\Psi}(\bar{\alpha}_{\ell_1\ell_2} + \bar{\beta}_{\ell_1\ell_2})) \\
	& + \sum_{j \neq i}\sum_{\ell_1, \ell_2 = 1}^k q_{j\ell_2} A_{ij} Z_{i\ell_1}  (\bm{\Psi}(\bar{\alpha}_{\ell_1\ell_2}) - \bm{\Psi}(\bar{\alpha}_{\ell_1\ell_2} + \bar{\beta}_{\ell_1\ell_2})) \\
	& + \sum_{j \neq i}\sum_{\ell_1, \ell_2 = 1}^k q_{j\ell_2} (1 - A_{ij}) Z_{i\ell_1} (\bm{\Psi}(\bar{\beta}_{\ell_1\ell_2}) - \bm{\Psi}(\bar{\alpha}_{\ell_1\ell_2} + \bar{\beta}_{\ell_1\ell_2})) \\
	&+ \textrm{const}_2.
\end{aligned}
\end{equation}
}%
As before, all terms that do not depend on $\bfz_i$ have been absorbed in $\textrm{const}_2$. We have also used the following two facts: \textbf{(i)} $X \sim \textrm{Beta}(\alpha, \beta) \Rightarrow \rmE[\ln X] = \bfPsi(\alpha) - \bfPsi(\alpha + \beta)$, and \textbf{(ii)} $X \sim \textrm{Beta}(\alpha, \beta) \Rightarrow 1 - X \sim \textrm{Beta}(\beta, \alpha)$. Using \eqref{eq:exp_log_joint_qn} in \eqref{eq:cavi_updates} provides the update equation for $q^*_i(\bfz_i)$.

Now that we have a way to update $\rmQ(\bfZ, \bfB)$ for a fixed $\bm{\theta}$ using \eqref{eq:cavi_updates}, we can proceed to the M-step.


\subsubsection{M-step}
\label{section:m_step}

In the M-step, the distribution $\rmQ$ is held constant and $\calL_{\rmQ}(\bm{\theta})$ is maximized over $\bfW, \bfu$, and $\bfv$. We do this by computing the gradient of $\calL_{\rmQ}(\bm{\theta})$ with respect to these parameters and performing gradient ascent. Using the structure of $\rmQ$ from \eqref{eq:mean_field} and linearity of expectation, we get
{
\small
\begin{equation}
\label{eq:expected_log_joint}
\begin{aligned}
	\calL_{\rmQ}(\bm{\theta}) &= \sum_{i=1}^{n} \Big(\sum_{j=1}^m \sum_{\ell=1}^k Y_{ij} q_{i}(\ell) W_{j\ell} + \sum_{j=1}^m u_j Y_{ij} \\
    &\;\;\; + \sum_{\ell=1}^k v_\ell q_i(\ell) \Big) - n \ln \Psi(\bfW, \bfu, \bfv) + \textrm{const}_3.
\end{aligned}
\end{equation}
}%
All terms not involving $\bfW, \bfu$, and $\bfv$ have been absorbed in $\textrm{const}_3$. Differentiating \eqref{eq:expected_log_joint} results in:
{
\small
\begin{equation}
\label{eq:gradients}
\begin{aligned}
	\nabla_{W_{j\ell}} \calL_{\rmQ}(\bm{\theta}) 
	&= \sum_{i=1}^n Y_{ij} q_i(\ell) -  n \rmE_{(\bfy, \bfz) \sim \rmP_{\mathrm{RBM}}}[y_j z_\ell]\Big], \\
    \nabla_{v_\ell} \calL_{\rmQ}(\bm{\theta}) &= \sum_{i=1}^n q_{i}(\ell) - n \rmE_{(\bfy, \bfz) \sim \rmP_{\mathrm{RBM}}}[z_\ell], \text{ and }\\
    \nabla_{u_j} \calL_{\rmQ}(\bm{\theta}) &= \sum_{i=1}^n Y_{ij} - n \rmE_{(\bfy, \bfz) \sim \rmP_{\mathrm{RBM}}}[y_j].
\end{aligned}
\end{equation}
}%
Above, $\rmP_{\mathrm{RBM}}$ denotes the current joint probability distribution over covariates and community memberships encoded by the RBM. These gradients are used for gradient ascent with a learning rate $\epsilon > 0$. The choice of learning rate is important, but there is a wide range of values for which the optimization procedure is stable and converges at a reasonable pace. We use $\epsilon=1/n$ in all experiments as the \textrm{ELBO} scales linearly with $n$. 

This completes the specification of the M-step. Figure \ref{alg:inference_procedure} summarizes the inference procedure. We next describe a few practical considerations.

\begin{figure}
\begin{algorithmic}[1]
    \STATE \textbf{Input:} $\bfA$, $\bfY$, $\bm{\alpha}$, $\bm{\beta}$, batch-size $b$, maximum iterations $\tau$, update steps $\xi$, and learning rate $\epsilon$
    \STATE Initialize $\rmQ$, $\bfW$, $\bfu$ and $\bfv$
    \FOR{$\tau$ iterations}
        \STATE \textbf{E-step:}
        \FOR{all pairs $i, j \in \{1, \dots, k\}$}
            \STATE Update $q^*_{ij}$ using \eqref{eq:cavi_updates}
        \ENDFOR
        \FOR{$b$ iterations}
            \STATE Select a random node $i$ from $\{1, \dots, n\}$
            \STATE Update $q^*_i$ using \eqref{eq:cavi_updates}
        \ENDFOR
        \STATE \textbf{M-step:} \\
        \FOR{$\xi$ iterations}
            \STATE Obtain gradients in \eqref{eq:gradients} and update $\bfW, \bfu$, and $\bfv$ using gradient ascent
        \ENDFOR
    \ENDFOR
\end{algorithmic}
\caption{Inference in \short{}}
\label{alg:inference_procedure}
\end{figure}

\paragraph{Batch updates in E-step} In a single E-step, we update $q^*_i$ only for a randomly chosen subset of $b$ nodes. This has a practical motivation, as smaller values of $b$ keep the computational burden of the E-step relatively low. However, $b$ should be large enough to ensure sufficient progress before the next M-step. Setting $b=\min\{n, 256\}$ works well in all our experiments, even when $n$ is as large as $100000$.

\paragraph{Gradient ascent in M-step} Ideally, one would run gradient ascent until convergence, to completely optimize $\calL_\rmQ(\bm{\theta})$ with respect to the RBM parameters at each M-step. However, we execute only $\xi=1$ gradient ascent steps in our experiments as done in \cite{Airoldi:2008:MixedMembershipStochasticBlockmodels}, and it works well in practice. Larger values of $\xi$ can be used, but empirically we show no signs of significant gains.

\paragraph{Exact computation of gradients in \eqref{eq:gradients}} The expectations in \eqref{eq:gradients} can be approximated using Monte-Carlo sampling, as is the standard practice for RBMs \cite{FischerEtAl:2012:AnIntroductionToRestrictedBoltzmannMachines}. However, because the normalization constant $\Psi(\bfW, \bfu, \bfv)$ can be efficiently computed in our case, we can also derive exact expressions for these terms (see Appendix \ref{appendix:exact_gradient_computation}). While the exact gradient computation is slightly faster, the gradients approximated via Gibbs sampling are observed to be numerically more stable.

\paragraph{Complexity of the inference procedure} In the E-step, we need $\bar{\alpha}_{ij}$ and $\bar{\beta}_{ij}$ for computing $q^*_{ij}$. These quantities can be computed in $O(n + \abs{\calE})$ steps, where $\calE$ is set of edges in the network, using the following reformulation.
{
\small
\begin{align*}
	\sum_{i_1 \neq i_2} (1 - A_{i_1i_2}) &q_{i_1}(i) q_{i_2}(j) =  \Big(\sum_{i_1 = 1}^n q_{i_1}(i)\Big) \Big(\sum_{i_2 = 1}^n q_{i_2}(j)\Big) \\
    &- \sum_{(i_1, i_2) \in \calE} q_{i_1}(i) q_{i_2}(j) - \sum_{i_1=1}^n q_{i_1}(i) q_{i_1}(j),
\end{align*}
}%
Thus, the total cost of updating $q^*_{ij}$ for all community pairs is $O(k^2(n + \abs{\calE}))$. 

Similarly, for computing $q^*_i$ using \eqref{eq:exp_log_joint_qn}, $\bfPsi(\bar{\alpha}_{\ell_1\ell_2}) - \bfPsi(\bar{\alpha}_{\ell_1\ell_2} + \bar{\beta}_{\ell_1\ell_2})$ and $\bfPsi(\bar{\beta}_{\ell_1\ell_2}) - \bfPsi(\bar{\alpha}_{\ell_1\ell_2} + \bar{\beta}_{\ell_1\ell_2})$ can be computed for all $\ell_1, \ell_2 = 1, \dots, k$ at the beginning of the E-step in $O(k^2)$ time. Using these quantities, $q^*_{i}$ can be computed for a fixed node $i$ in $O(n(m + k))$ steps. Because the inference procedure chooses to update $b$ nodes at each E-step, the total complexity of an E-step is given by $O(k^2(n + \abs{\calE}) + k^2 + bn(m + k))$. The computation over all community pairs and attributes can be done in parallel and $b$ is a fixed constant. Hence, the effective time needed is $O(n + \abs{\calE})$.

In the M-step, the time needed for computing the expectations in \eqref{eq:gradients} is $O(mk)$, which is independent of $n$ and $\abs{\calE}$. However, \eqref{eq:gradients} involves a summation over all nodes. Thus, an M-step takes $O(nmk)$ time. As before, the computation over communities and attributes can be parallelized, yielding an effective running time of $O(n)$. 

Therefore, each step in parallel implementation of the inference procedure runs in time $O(n + \abs{\calE})$, making it suitable for large-scale applications. We run the inference for a fixed number of $\tau = 1000$ iterations in our experiments. One can also use other stopping criteria like a minimum improvement in the value of \textrm{ELBO}.

In the next section we relax the assumption of binary covariates.


\subsection{\short{} for Continuous Covariates}
\label{section:continuous_rbsbm}

This section describes a variant of \short{} that allows the covariates to take continuous values. We will assume that the covariates are bounded and scale them to lie in the range $[0, 1]$, thus $\bfY \in [0, 1]^{n \times m}$. The RBM must be modified to accommodate continuous valued inputs. First, the normalization constant $\Psi(\bfW, \bfu, \bfv)$ changes to reflect the fact that $\bfy \in [0, 1]^m$. Next, the conditional distribution $\rmP_{\bm{\theta}}(\bfy \vert \bfz)$ changes to
{
\small
\begin{align*}
	\label{eq:prob_y_given_z_cont}
	&\rmP_{\bm{\theta}}(y_j = y \vert z_\ell = 1) = \rmP_{\bm{\theta}}(y_j = y \vert z_\ell = 1, \bfy_{-j}) \\
	&= \frac{\exp\Big((W_{j\ell} + u_j)y + \sum_{j^{'} \neq j} (W_{j^{'}\ell} + u_{j^{'}}) y_{j^{'}} + v_\ell\Big)}{\int_0^1 \exp\Big((W_{j\ell} + u_j)\bar{y} + \sum_{j^{'} \neq j} (W_{j^{'}\ell} + u_{j^{'}}) y_{j^{'}} + v_\ell\Big) \mathrm{d}\bar{y}} \\
    &= \frac{W_{j\ell} + u_j}{\exp(W_{j\ell} + u_j) - 1} \exp((W_{j\ell} + u_j) y).
\end{align*}
}%
The vector $\bfy_{-j} \in [0, 1]^{m - 1}$ has all entries of $\bfy$ except the entry at $j^{th}$ position. Samples can be drawn from the distribution above using inverse sampling. The form of $\rmP_{\bm{\theta}}(\bfz \vert \bfy)$ remains the same as before, except that the entries of $\bfy$ can now lie in $[0, 1]$. Other parts of the model remain unchanged.

As before, we use variational inference. Looking at the inference procedure in Section \ref{section:inference_in_rbsbm}, one sees that the binary assumption on the attributes plays a role only during the M-step, namely while using Gibbs sampling to compute the gradients. Therefore, the same inference procedure applies here as well, except for Gibbs sampling, which must be performed using the modified conditional distribution for the RBM given above. We experiment with this variant of \short{} in Section \ref{section:experiments}.


\section{\shortmm}
\label{section:rbmmsbm}

\short{} is a flexible model that can represent a large class of networks. However, it assumes that each node belongs to exactly one community. This assumption is violated in many practical instances, such as a collaboration network where a researcher can be active in many fields. In this section, we present our second model, \shortmm{}, that addresses this shortcoming by supporting mixed-community memberships. Sections \ref{section:rbmmsbm_description}  and \ref{section:inference_in_rbmmsbm} describe the model and the corresponding inference procedure, respectively. For simplicity, we assume that the covariates are binary, but the model can also support continuous covariates using a strategy similar to the one described in Section \ref{section:continuous_rbsbm}.


\subsection{Model Description}
\label{section:rbmmsbm_description}

As a motivational example, consider a collaboration network where nodes represent computer scientists and edges are present if two scientists collaborated on a project. One community structure of interest in that characterized by research area, such as machine learning, database systems, computer architecture, compilers, and so on. It is very restrictive to assume that a scientist can contribute to only one area. Mixed-membership allows the nodes to belong to multiple communities simultaneously. For example, a person may be knowledgeable in databases while having expertise in compilers. Most of their collaborations with others will be as an expert in compilers, however, they may also get involved in database-related projects. Following \cite{Airoldi:2008:MixedMembershipStochasticBlockmodels}, we assume that each node has a mixed-membership vector $\bfz_i \in \Delta_k$, where the $j^{th}$ entry captures the expected proportion of interactions in which node $i$ acts as a member of community $j$. Contrast this with the previous case where $\bfz_i$ was one-hot encoded, and it was assumed that node $i$ has the same behavior in all its interactions.

Similar to \short{}, \shortmm{} integrates node covariates into the well-known mixed-membership SBM through a variant of RBM. First, we describe the proposed changes to the RBM to accommodate mixed-membership vectors $\bfz_i$ that now belong to a simplex $\Delta_k$. Then, we sample $\bfz_i$'s from the RBM as a function of node covariates and use them to sample the edges in the network via a mixed-membership SBM. 

The joint distribution encoded by the RBM has the form
\begin{equation}
    \label{eq:prob_y_z_rbmmsbm}
    \rmP_{\bm{\theta}}(\bfy, \bfz) = \frac{\exp(\bfy^\intercal \bfW \bfz + \bfy^\intercal \bfu + \bfz^\intercal \bfv)}{\Omega(\bfW, \bfu, \bfv)},
\end{equation}
where $\Omega(\bfW, \bfv, \bfu)$ is the normalization constant given by
{
\small
\begin{equation}
    \label{eq:normalization_constant_rbmmsbm}
    \Omega(\bfW, \bfv, \bfu) = \sum_{\bfy \in \{0, 1\}^m} \int_{\bfz \in \Delta_k} \exp(\bfy^\intercal \bfW \bfz + \bfy^\intercal \bfu + \bfz^\intercal \bfv) \mathrm{d}\bfz.
\end{equation}
}%
As before, we use the following conditional distributions to draw samples from \eqref{eq:prob_y_z_rbmmsbm} via Gibbs sampling.
{
\small
\begin{equation}
\label{eq:conditional_distributions_rbmmsbm}
\begin{aligned}
    \rmP_{\bm{\theta}}(y_j = 1 | \bfz) &=  \sigma(\sum_{\ell=1}^k W_{j\ell} z_\ell + u_j) \text{ and} \\
    \rmP_{\bm{\theta}}(z_\ell = z, z_k = 1 - z - &s_\ell \,\vert\, \bfy, \bar{\bfz}_{\ell}) = \frac{\beta_{\ell} \exp(a \beta_\ell)}{\exp((1 - s_\ell)\beta_\ell) - 1}.
\end{aligned}
\end{equation}
}%
Here, $\sigma(x) = \frac{1}{1 + \exp(-x)}$ is the logistic sigmoid function, $\beta_\ell = v_\ell - v_k + \sum_{j=1}^m (W_{j\ell} - W_{jk})y_j$, $\bar{\bfz}_{\ell}$ is a vector containing all elements of $\bfz$ except the $\ell^{th}$ and $k^{th}$ elements, and $s_\ell$ is the sum of all entries in $\bar{\bfz}_{\ell}$. The derivation of \eqref{eq:conditional_distributions_rbmmsbm} is given in Appendix \ref{appendix:sampling_from_modified_rbm_mmsbm}.

Next, we use the mixed-membership SBM to model the edges in the network. Let $\bfpsi^{(i)}_{ij} \in \bbR^k$ be a one-hot encoded vector that specifies the role assumed by node $i$ (the community to which it belongs) for a directed interaction from node $i$ to node $j$. These \textit{interaction specific} vectors $\bfpsi^{(i)}_{ij}$ allow nodes to assume different community memberships while interacting with different nodes. To model a directed edge from node $i$ to node $j$, first we sample the community memberships from $\bfz_i$ and $\bfz_j$ to get $\bfpsi^{(i)}_{ij}$ and $\bfpsi^{(j)}_{ij}$, respectively. Then, the probability of the edge is given by $\rmP(A_{ij} = 1 \vert \bfpsi^{(i)}_{ij}, \bfpsi^{(j)}_{ij}, \bfB) = {\bfpsi^{(i)}_{ij}}^\intercal \bfB \bfpsi^{(j)}_{ij}$, where $\bfB \in [0, 1]^{k \times k}$ is a block matrix encoding the relationship between various communities, as in \short{}. While the traditional mixed-membership SBM samples the vectors $\bfz_i$ from a Dirichlet distribution, we sample them from the RBM. As before, we follow a Bayesian approach and assume that $B_{ij} \sim \textrm{Beta}(\alpha_{ij}, \beta_{ij})$ for all entries in $\bfB$, where $\alpha_{ij}, \beta_{ij} > 0$ are user specified hyperparameters.

The process of generating networks from \shortmm{} is summarized below.
\begin{enumerate}
	\item  Sample $B_{ij} \sim \textrm{Beta}(\alpha_{ij}, \beta_{ij})$ for all $i, j = 1, \dots, k$.
	\item Sample $(\bfy_i, \bfz_i) \sim \rmP_{\bm{\theta}}(\bfy, \bfz)$ for all $i = 1, \dots, n$ from the RBM.
	\item  For each pair of nodes $i \neq j$,
	\begin{enumerate}
		\item Sample $\bfpsi^{(i)}_{ij} \sim \bfz_i$
		\item Sample $\bfpsi^{(j)}_{ij} \sim \bfz_j$
		\item Sample $A_{ij} \sim \textrm{Bernoulli}({\bfpsi^{(i)}_{ij}}^\intercal \bfB \bfpsi^{(j)}_{ij})$.
	\end{enumerate}
\end{enumerate}
\shortmm{} inherits several desirable properties from \short{} while being more general. For example, as before, it can be used to explain the communities, and it is not domain-specific. In fact, the model can provide additional insights about the roles played by the nodes for each interaction in the network through the vectors $\bfpsi^{(i)}_{ij}$.

Let $\bfPsi^+ = \{\bfpsi^{(i)}_{ij}: i, j = 1, \dots, n\}$ and $\bfPsi^- = \{\bfpsi^{(j)}_{ij}: i, j = 1, \dots, n\}$, and define $\bfPsi = \bfPsi^+ \cup \bfPsi^-$. Using the generative process specified above, the joint log-probability is given by
{
\small
\begin{equation*}
    \label{eq:joint_distr_compressed_mmrbsbm}
    \rmP_{\bm{\theta}}(\bfA, \bfY, \bfZ, \bfPsi, \bfB) = \rmP_{\bm{\theta}}(\bfB) \, \rmP_{\bm{\theta}}(\bfY, \bfZ) \, \rmP_{\bm{\theta}}(\bfPsi \vert \bfZ) \, \rmP_{\bm{\theta}}(\bfA \vert \bfB, \bfPsi).
\end{equation*}
}
Next, we derive the inference procedure for \shortmm{}.


\subsection{Inference in \shortmm{}}
\label{section:inference_in_rbmmsbm}

Inference in \shortmm{} follows the same recipe as before, with one key exception. We begin by identifying the latent variables and parameters and compute the \textrm{ELBO}. However, unlike the last time, it is no longer possible to derive a closed form solution for the optimal factors in the approximating distribution $\rmQ$ (more details are given below). To address this issue, we make further assumptions about the parametric form of $\rmQ$ and use gradient ascent to maximize \textrm{ELBO} with respect to both $\rmQ$ and the model parameters. 

The latent random variables in \shortmm{} are the mixed-membership vectors $\bfZ$, the edge-specific membership indicators $\bfPsi$, and the block matrix $\bfB$. In addition, \textrm{ELBO} also depends on RBM parameters $\bfW$, $\bfu$, and $\bfv$. As before, we observe the adjacency matrix $\bfA$ and the node covariates $\bfY$ and use variational inference. The \textrm{ELBO}, denoted by $\calL_{\rmQ}(\bm{\theta})$, is given by, 
{
\small
\begin{equation}
\label{eq:elbo_mmrbsbm}
    \calL_{\rmQ}(\bm{\theta}) = \rmE_{\rmQ}[\ln \rmP_{\bm{\theta}}(\bfA, \bfY, \bfZ, \bfPsi, \bfB)] -  \rmE_{\rmQ}[\ln \rmQ(\bfZ, \bfPsi,\bfB)]
\end{equation}
}%
where $\rmQ(\bfZ, \bfPsi,\bfB)$ approximates the true posterior over the latent random variables. We again assume that $\rmQ$ belongs to the mean-field family of distributions. Thus,
{
\small
\begin{equation}
\label{eq:mean_field_mmsbm}
\begin{aligned}
    \rmQ(\bfZ, \bfPsi,\bfB) =& \Big(\prod_{i=1}^n q_i(\bfz_i) \Big) \Big(
    \prod_{i \neq j} q_{ij}^{(i)}(\bfpsi^{(i)}_{ij})\Big) \\
    &\Big(\prod_{i \neq j} q_{ij}^{(j)}(\bfpsi^{(j)}_{ij})\Big)
    \Big(\prod_{i, j = 1}^k q_{ij}(B_{ij})\Big).
\end{aligned}
\end{equation}
}
This time, deriving update equations for the factors in the expression above (as in \eqref{eq:cavi_updates}) is not straightforward. Notice that \eqref{eq:cavi_updates} requires one to compute certain expectations inside the exponential function. For \shortmm{}, these expectations are hard to compute as they require evaluating integrals over a simplex $\Delta_k$ for the terms containing $\bfZ$. In the standard mixed-membership SBM \cite{Airoldi:2008:MixedMembershipStochasticBlockmodels}, the authors assume a Dirichlet prior on $\bfZ$. This makes the computation of expectations easy, as the Dirichlet distribution is the conjugate prior for the multinomial distribution. In our case, $\bfZ$ is sampled from an RBM. Nonetheless, we approximate it by assuming that $q_i(\bfz_i)$ is a Dirichlet distribution to make the calculations tractable. More precisely, we make the following assumptions about the parametric form of various factors in \eqref{eq:mean_field_mmsbm}:
\begin{enumerate}
    \item $q_i(\bfz_i)$ is a Dirichlet distribution with parameter $\bfmu_i = [\mu_{i1}, \mu_{i2}, \dots, \mu_{ik}]$.
    \item $q_{ij}^{(i)}(\bfpsi^{(i)}_{ij} = \bfe_\ell) = \alpha^{(i)}_{ij}(\ell) \in [0, 1]$, where $\bfe_\ell$ is the $\ell^{th}$ one-hot encoded vector and $\alpha^{(i)}_{ij}(\ell)$ is a parameter. Similarly, $q_{ij}^{(j)}(\bfpsi^{(j)}_{ij})$ is also a multinomial distribution such that $q_{ij}^{(j)}(\bfpsi^{(j)}_{ij} = \bfe_\ell) = \alpha^{(j)}_{ij}(\ell) \in [0, 1]$.
    \item $q_{ij}(B_{ij})$ is a Beta distribution with parameters $\bar{\alpha}_{ij}$ and $\bar{\beta}_{ij}.$
\end{enumerate}
We will also use $q_i(\ell)$ to denote $\rmE_{\bfz \sim q_i}[z_{\ell}]$, thus, $q_i(\ell) = \frac{\mu_{i\ell}}{\sum_{j = 1}^k \mu_{ij}}$. These assumptions allow us to compute \textrm{ELBO}. See Appendix \ref{appendix:rbmmsbm_elbo_derivation} for details. The optimization of \textrm{ELBO} is again divided into two steps:
\begin{enumerate}
    \item \textbf{E-step}: Optimize over the parameters of the factors in $\rmQ$ keeping $\bfW$, $\bfu$, and $\bfv$ fixed.
    \item \textbf{M-step}: Optimize over the parameters $\bfW$, $\bfu$, and $\bfv$ for a fixed $\rmQ$.
\end{enumerate}
In both the steps above, we compute the derivative of \textrm{ELBO} with respect to the corresponding parameters, and approximately maximize \textrm{ELBO} using gradient ascent. The expressions for gradients with respect to $\bfW$, $\bfu$, and $\bfv$ are same as \eqref{eq:gradients} (using $q_i(\ell) = \frac{\mu_{i\ell}}{\sum_{j = 1}^k \mu_{ij}}$). We used PyTorch's automatic differentiation \cite{PaszkeEtAl:2019:PyTorch} to compute derivatives with respect to other parameters in our experiments, though one can try to compute these values manually as well.

\begin{figure}
\begin{algorithmic}[1]
    \STATE \textbf{Input:} $\bfA$, $\bfY$, $\bm{\alpha}$, $\bm{\beta}$, maximum iterations $\tau$, update steps $\xi$, and learning rate $\epsilon$
    \STATE Initialize $\rmQ$, $\bfW$, $\bfu$ and $\bfv$
    \FOR {$\tau$ iterations}
        \STATE \textbf{E-step:}
        \FOR{$\xi$ iterations}
            \STATE Compute derivative of \textrm{ELBO} with respect to parameters in $\rmQ$
            \STATE Update the parameters of factors in $\rmQ$ using gradient ascent while respecting the constraints (see Section \ref{section:inference_in_rbmmsbm})
        \ENDFOR
        \STATE \textbf{M-step:}
        \FOR{$\xi$ iterations}
            \STATE Obtain gradients with respect to $\bfW, \bfu$, and $\bfv$ and update these parameters
        \ENDFOR
    \ENDFOR
\end{algorithmic}
\caption{Inference in \shortmm{}}
\label{alg:inference_procedure_rbmmsbm}
\end{figure}

Figure \ref{alg:inference_procedure_rbmmsbm} summarizes the inference algorithm. Next, we make a few remarks about the practical implementation.

\paragraph{Constrained optimization during E-step} The parameters in $\rmQ$ cannot take arbitrary values. For example, $\sum_{\ell=1}^k \alpha^{(i)}_{ij}(\ell) = 1$ and all parameters must be non-negative. One approach is to add these constraints to the optimization problem. However, constrained optimization problems are harder to solve. A simpler workaround is to reparameterize these quantities so that they always satisfy the constraints by construction. For example, we use a softmax function to produce $[\alpha^{(i)}_{ij}(1), \dots, \alpha^{(i)}_{ij}(k)]$ and optimize the input to this softmax function instead of directly optimizing $[\alpha^{(i)}_{ij}(1), \dots, \alpha^{(i)}_{ij}(k)]$. This is easy to do because the input to the softmax function is unconstrained. We similarly reparameterize other variables by using the exponential function to ensure that they are always positive.

\paragraph{Computing gradients during M-step} Unlike \short{}, the normalization constant for RBM cannot be computed exactly in this case. Therefore, to compute the gradients with respect to $\bfW$, $\bfu$, and $\bfv$, we approximate the expectation terms in \eqref{eq:gradients} using Monte-Carlo estimation. Following the standard practice for RBMs \cite{FischerEtAl:2012:AnIntroductionToRestrictedBoltzmannMachines}, we use $\eta$ persistent Gibbs chains to sample from the RBM using \eqref{eq:conditional_distributions_rbmmsbm}. These samples are then used to approximate the required expectations in \eqref{eq:gradients}.

\paragraph{Not running E and M-steps until convergence} As in the previous case, we do not run the M-step until convergence but instead take only $\xi = 1$ gradient ascent steps. Similarly, because the E-step is also gradient ascent based for \shortmm{}, we only take $\xi=1$ gradient steps. Empirically this performs essentially as well as using multiple gradient steps in both the E and M steps.

\paragraph{Computational complexity} The complexity is dominated by the calculation of \textrm{ELBO} in E-step, which takes $O(n^2)$ operations. This quadratic dependence on $n$ limits the size of networks that we can experiment with. However, it must be noted that the standard mixed-membership SBM has the same time complexity, and improving upon it is not possible as there are $O(n^2)$ elements in $\bfPsi$ to be inferred. Nonetheless, the method can be easily applied to networks with a few thousand nodes, which is still useful in many practical instances, especially because the model offers additional information about interaction-specific roles ($\bfPsi$), which is not available in models with pure community memberships.


\section{Experiments}
\label{section:experiments}

This section is divided into four parts. Section \ref{section:datasets} describes the synthetic and real-world networks used in our experiments. Section \ref{section:implementation_details} contains practical implementation details. Finally, Sections \ref{section:experiments_rbsbm} and \ref{section:experiments_rbmmsbm} describe the experimental results related to \short{} and \shortmm{}, respectively.\footnote{Code: \url{https://github.com/sml-iisc/NetworksWithAttributes}}


\subsection{Datasets}
\label{section:datasets}

We use the following real-world networks.

\paragraph{Cora} This is a citation network where nodes represent publications, and directed edges represent the ``cites'' relationship \cite{LuGetoor:2003:LinkBasedClassification}. There are $2708$ nodes and $5429$ edges in this network. Each node also has a $1433$-dimensional binary covariate vector whose elements indicate the absence or presence of various words in the publication. Seven ground-truth communities are known.

\paragraph{Citeseer} This is also a citation network similar to Cora \cite{LuGetoor:2003:LinkBasedClassification}. It has $3312$ nodes, $4732$ edges, $3703$-dimensional binary covariates, and $6$ ground-truth communities.

\paragraph{Philosophers} This network was crawled from Wikipedia (see also \cite{YangEtAl:2013:CommunityDetectionInNetworksWithNodeAttributes}). Nodes represent philosophers listed on Wikipedia\footnote{\url{https://en.wikipedia.org/wiki/Lists_of_philosophers}}. A directed edge from node $i$ to node $j$ indicates that the Wikipedia page of node $i$ links to the Wikipedia page of node $j$. Covariates indicate the presence of links to other non-philosopher entries on Wikipedia. We discard covariates for which the value is one for less than ten or more than $50\%$ of the nodes. This network has $1497$ nodes, $44996$ edges, and $6357$ covariates. The ground-truth communities are unknown.

\paragraph{Sinanet} This network has $3490$ nodes representing users on a micro-blogging website, $30282$ edges encoding the ``follows'' relation between them and $10$-dimensional continuous node covariates. The covariate vectors belong to a simplex and represent the proportion of a user's interest in various topics. Ten ground-truth communities are known which correspond to various forums on the website \cite{JiaEtAl:2017:NodeAttributeEnhancedCommunityDetectionInNetworks}.

\paragraph{PubMed Diabetes} This is also a citation network. Nodes represent publications on Diabetes from the PubMed database. There are $19717$ nodes and $44338$ edges. Each node has a $500$-dimensional continuous covariate vector describing the TF/IDF weight for different words in the publication \cite{NamataEtAl:2012:QueryDrivenActiveSurveyingForCollectiveClassification}. We refer to this dataset as PubMed.

\paragraph{Lazega Lawyers} There are $71$ nodes in this network corresponding to layers at a firm. Edges encode friendships between them ($\text{\#edges} = 575$). Nodes have attributes like gender, age, law school attended, and so on. We discretize the value of these attributes to get $24$ binary covariates, as described in Appendix \ref{appendix:experiments}.

Recall that both \short{} and \shortmm{} are generative models. In addition to the real-world networks, we also use the following synthetic networks generated using our models.

\paragraph{Synth-$n$} Synth-$n$ is a $n$ node network sampled from \short{} (see Section \ref{section:rbsbm_description}). We set $\alpha_{ij}=1$ for all $i, j = 1, \dots, k$ and $\beta_{ij} = \sqrt{n}$ if $i=j$ and $10\sqrt{n}$ otherwise. This choice roughly implies that the sparsity of sampled networks is $O(1 / \sqrt{n})$ and edges within communities are $10$ times more likely than the edges across communities. To generate covariates, we assume that each covariate has an assortative role (nodes with same covariate value are more likely in the same community), disassortative role, or neutral role. In each community $\ell$, each covariate $j$ is assigned one of these roles with probabilities $p_+ = 0.1$, $p_- = 0.1$, and $p_0 = 0.8$, respectively, and the value of $W_{j\ell}$ is set to $+5$, $-5$, and $0$, respectively. All elements of $\bfu$ are set to $-2$ and all elements of $\bfv$ are set to $0$. In these networks, we use $m = 100$ and $k = \log_2 n$.

\paragraph{SynthMM-n-m-k} These networks exhibit mixed community memberships and are sampled from \shortmm{} (see Section \ref{section:rbmmsbm_description}). We set $\alpha_{ij}$, $\beta_{ij}$, $\bfW$, $\bfu$, and $\bfv$, as in Synth-$n$ above. The values of $n$, $m$, and $k$ are specified during sampling.


\subsection{Implementation details}
\label{section:implementation_details}

This section presents details about our implementation and some practical tricks that were found to improve the stability of the inference procedure.

\paragraph{Initialization} Both $q_{i}$ and $q_{ij}$ are updated during the E-step in Algorithm \ref{alg:inference_procedure}. The order in which these updates are made is important from the numerical stability perspective. A bad initialization of $q_{ij}$ leads to numerical overflows while computing the exponential in \eqref{eq:cavi_updates} for updating $q_i$. Thus, we update $q_{ij}$ first for all $i, j = 1, \dots, k$ before updating any of the $q_i$'s. This means that we only need to consider initialization for $q_i$'s. We simply set $q_i(\ell) = \frac{1}{k}$ for all nodes $i = 1, \dots, n$ and all communities $\ell = 1, \dots, k$.

\paragraph{Computing gradients in \eqref{eq:gradients}} We use exact computation of gradients (Appendix \ref{appendix:exact_gradient_computation}) while experimenting with Synth-$n$. In all other experiments, we approximate the gradients via Monte-Carlo estimation. For this, we use $\eta = 100$ persistent Gibbs chains and accept every tenth sample from these chains. The samples are then used to empirically approximate the expectations in \eqref{eq:gradients}.

\paragraph{Simulated annealing} If $q_i(\ell)$ becomes very small for all nodes $i = 1, \dots, n$ for a particular community $\ell$ during the initial few steps, the community $\ell$ effectively \textit{dies} (i.e., no nodes belong to this community). We observed that: \textbf{(i)} this is a common occurrence due to numerical issues at the beginning of the inference procedure when most quantities are far away from their optimal values, and \textbf{(ii)} a \textit{dead} community never comes back to life. To avoid this, we use a simulated annealing heuristic. After E-step, we apply the following transformation:
\begin{equation}
    \label{eq:h_transformation}
    h(x) = 
    \begin{cases} 
        \frac{1}{2^{\lambda - 1}} x^\lambda & \mbox{if } 0 \leq x \leq \frac{1}{2} \\ 
        1 - \frac{1}{2^{\lambda - 1}} (1-x)^\lambda& \mbox{if } \frac{1}{2} < x \leq 1
    \end{cases},
\end{equation}
to all $q_i(\ell)$'s that were updated during the E-step. At $\lambda=1$, $h(x)=x$ and for $\lambda < 1$, $h(x) > x$ if $x \leq 1/2$ and $h(x) < x$ otherwise. Thus, for $\lambda < 1$, this transformation dampens high values (close to 1) and elevates low values (close to 0) therefore achieving the regularization effect that prevents $q_i(\ell)$ from becoming too small for a particular community. Note that $q_i$ has to be re-normalized after applying the transformation $h(\cdot)$ so that its entries add up to one. We start with $\lambda = 0.3$ and increase it linearly to $1$ as the number of iterations increases.

\paragraph{Other hyper-parameters} We set batch-size $b = \min\{n, 256\}$, update steps $\xi = 1$, learning rate $\epsilon = 1/n$, and maximum number of iterations $\tau = 1000$, in all cases. We have justified the choice of these values in the previous sections. Our methods are relatively robust to the values of $b$, $\xi$, and $\epsilon$, as long as they are within reasonable limits. We set $\tau = 1000$ because $\textrm{ELBO}$ saturates before $1000$ steps in most of our experiments (except when very large synthetic networks are used). One may also use other criteria such as a minimum change in $\textrm{ELBO}$ per iteration to terminate the inference procedure. In our experiments with real data, we use $\alpha_{ij} = 1$ and $\beta_{ij} = 1$ if $i = j$ and $\beta_{ij} = 10$ otherwise for all $i, j = 1, \dots, k$. This prior guides the model towards discovering assortative communities. Other values can also be used to, for example, discover hierarchical communities, if such information is available apriori.


\subsection{Results for \short{}}
\label{section:experiments_rbsbm}

This section outlines several experiments involving the \short{} model. These show that the inference procedures scale well with the size of the networks and lead to meaningful inference results, for both community detection and for link prediction tasks. Furthermore, the inference results provide interpretable insights that help explaining the inference results. Such a feature is not typically available in many other methods with similar quantitative performance.


\begin{figure}
    \centering
    \subfloat[][]{\includegraphics[width=0.5\linewidth]{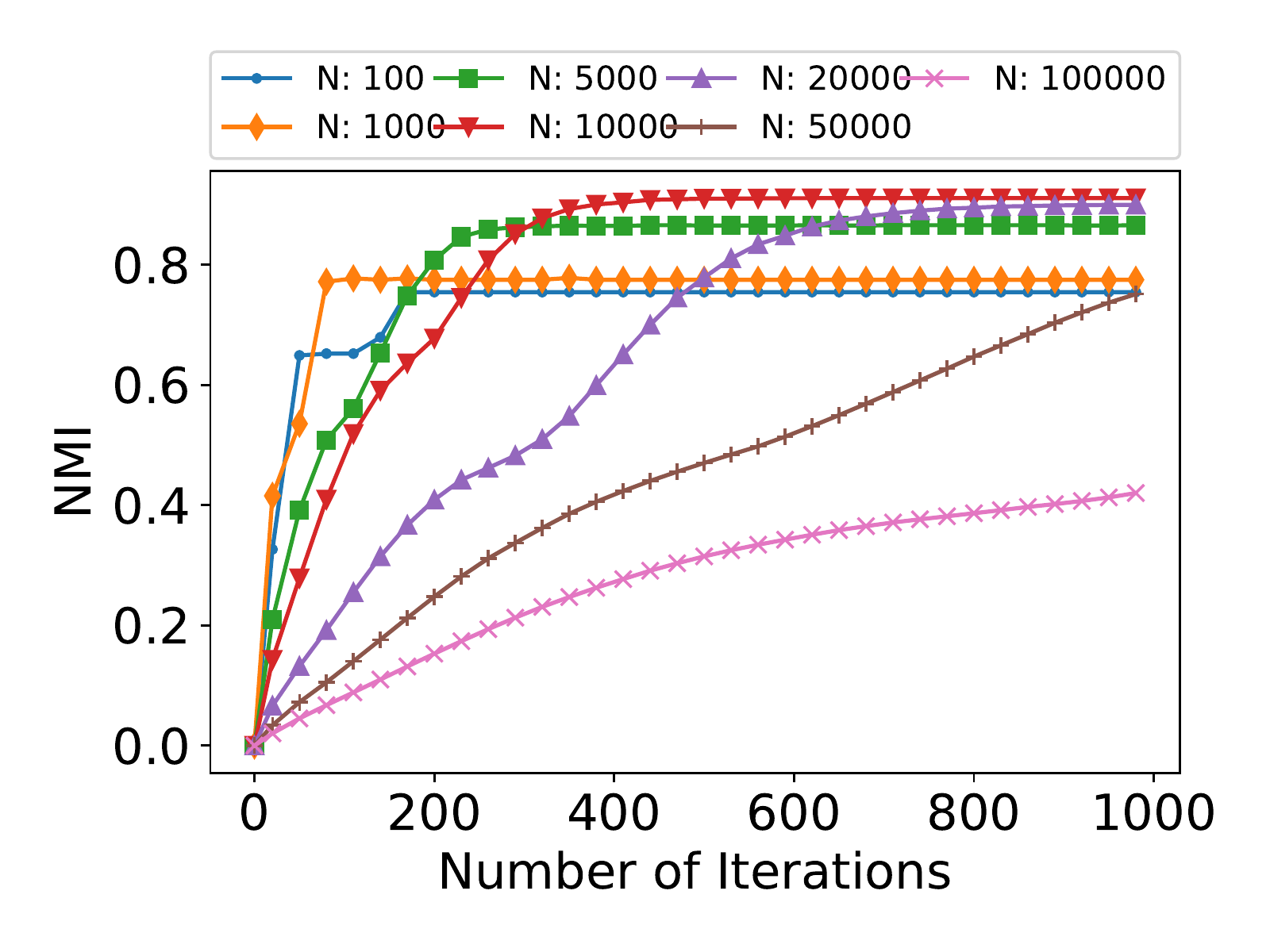}\label{fig:nmi_vs_iters}}%
    \subfloat[][]{\includegraphics[width=0.5\linewidth]{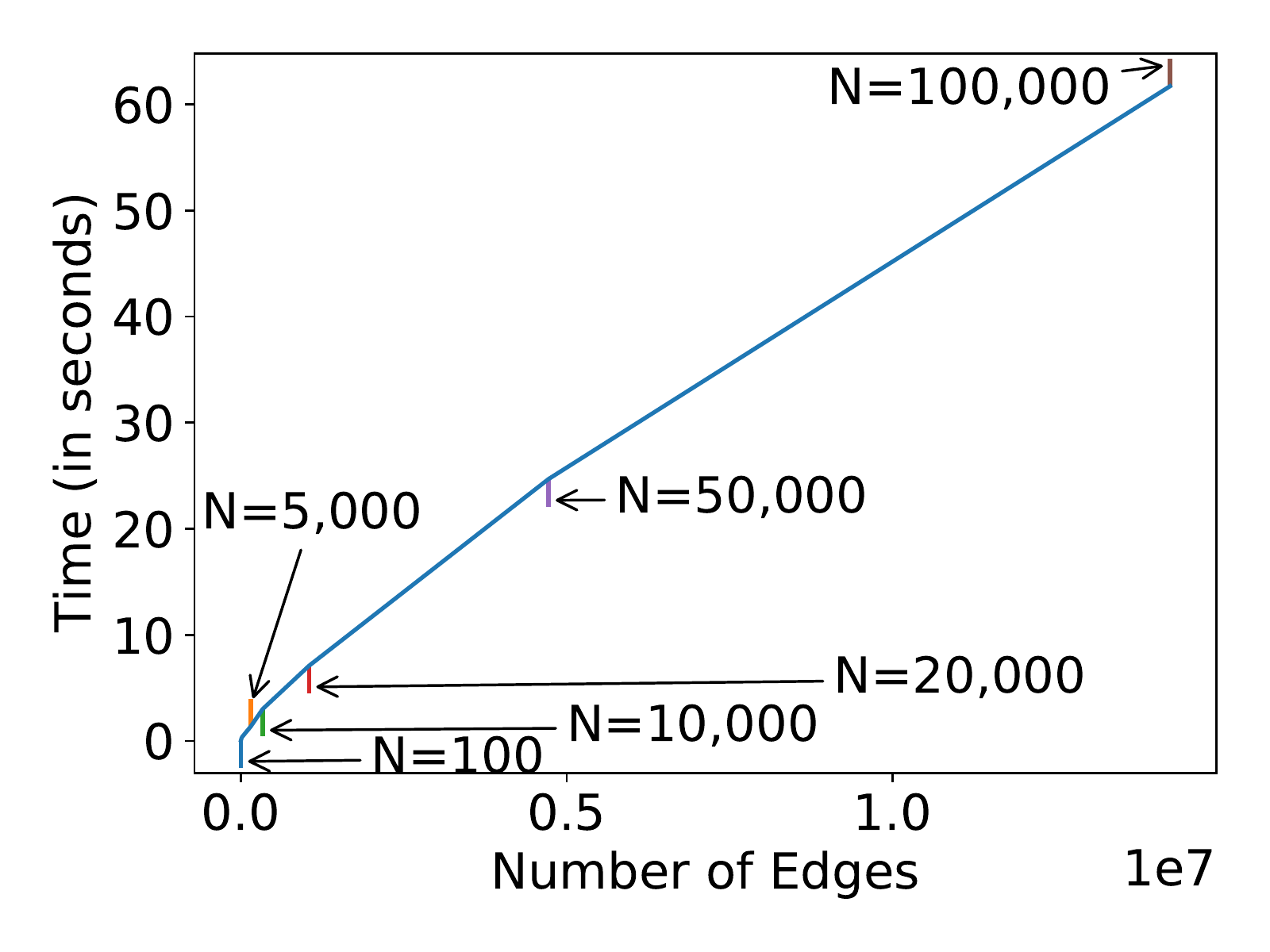}\label{fig:running_time}}
    \caption{Correctness and scalability of the inference procedure in \short{}. Panel~\protect\subref{fig:nmi_vs_iters} shows that NMI between discovered and ground-truth communities improves over time. Panel~\protect\subref{fig:running_time} shows that the average time taken for one iteration of E and M steps increases only linearly with the number of edges. We used Synth-$n$ networks for this experiment.}
    \label{fig:rbsbm_correctness_scalability}
\end{figure}

\paragraph{Correctness and scalability} We sample synthetic networks Synth-$n$ from \short{} for various values of $n$ and try to recover the communities from the observed network $\bfA$ and covariates $\bfY$ using the algorithm in Figure \ref{alg:inference_procedure}. Fig.~\ref{fig:nmi_vs_iters} shows the Normalized Mutual Information (NMI) scores \cite{DanonEtAl:2005:ComparingCommunityStructureIdentification} between detected community memberships and sampled ground-truth communities as a function of the number of iterations for the first $1000$ iterations. NMI takes values between 0 and 1, and larger values are more desirable. As expected, NMI increases over time, implying the correctness of the inference procedure. Since the batch size $b= \min\{256, n\}$, when $n$ is large, NMI increases steadily but at a slower pace. For example, when $n=100000$, after $1000$ iterations, each node's posterior over community membership has been updated less than three times on an average. After $5000$ iterations, the average NMI scores were $0.90$ and $0.76$ when $n=50000$ and $n = 100000$, respectively. Fig. ~\ref{fig:running_time} shows the average time taken for one iteration of E and M steps as a function of the number of edges. It also indicates the number of nodes for each data point. The running time scales linearly with the number of edges, as expected. This experiment was executed on an Intel Core $i7$-$6700$ machine with $4$ GB of main memory. For the continuous variant of \short{} our empirical results are rather similar and are not presented in this document.

\begin{figure*}
	\subfloat[][Members]{\includegraphics[width=0.24\linewidth,trim={3.6cm 1.3cm 3.2cm 1.3cm},clip]{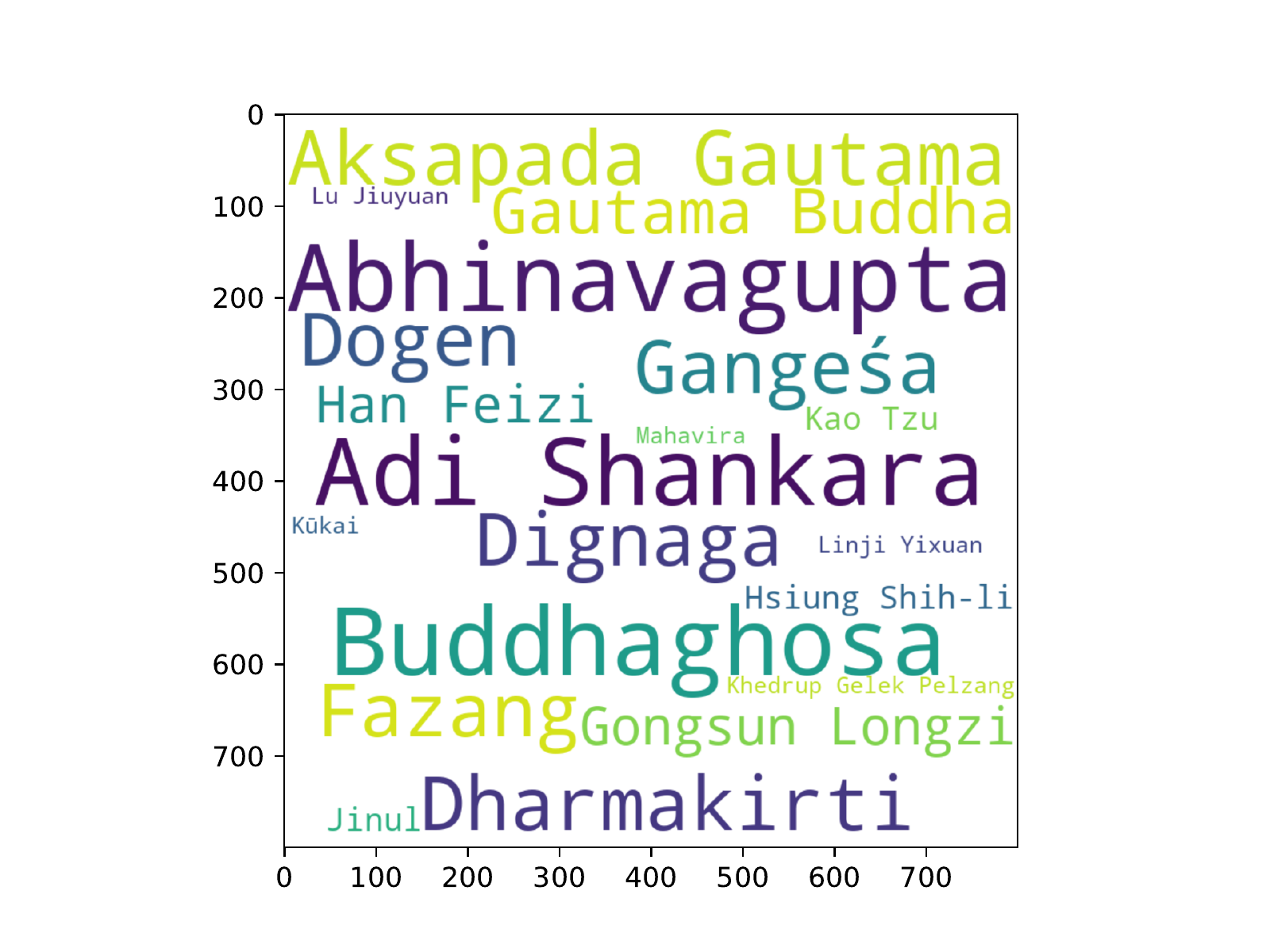}}%
	\hfill\subfloat[][Covariates]{\includegraphics[width=0.24\linewidth,trim={3.6cm 1.3cm 3.2cm 1.3cm},clip]{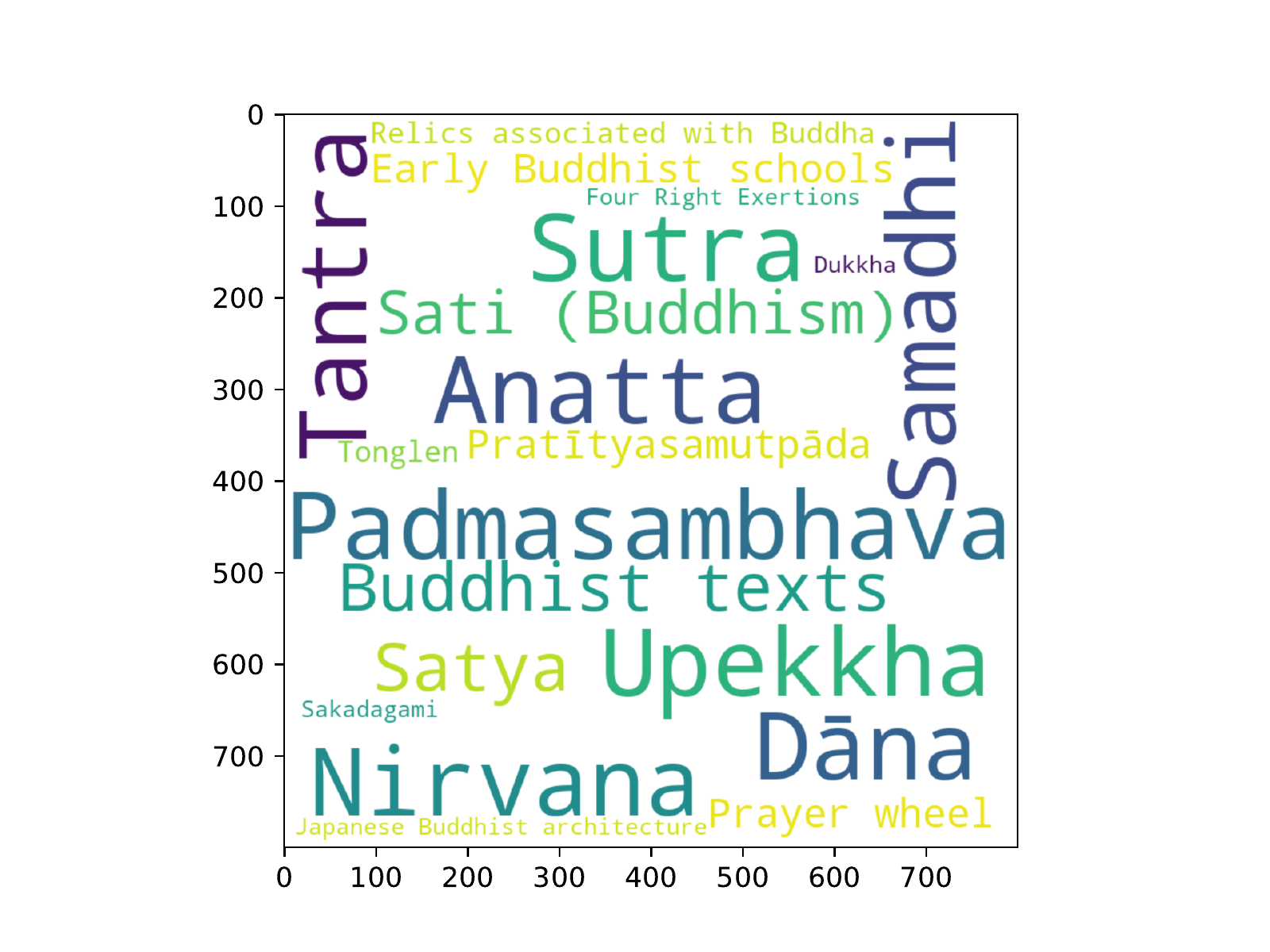}}%
	\hfill\subfloat[][Members]{\includegraphics[width=0.24\linewidth,trim={3.6cm 1.3cm 3.2cm 1.3cm},clip]{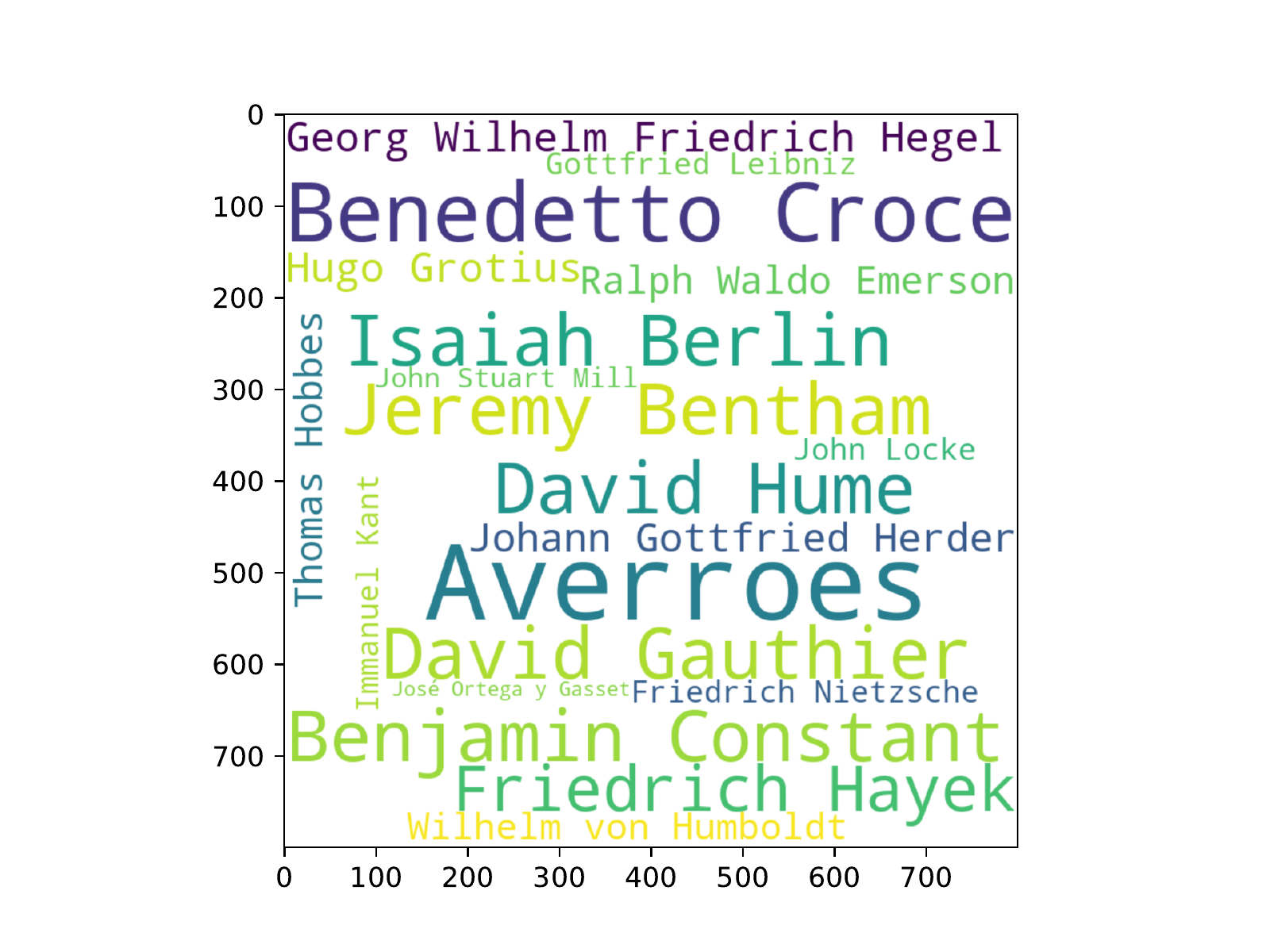}}%
	\hfill\subfloat[][Covariates]{\includegraphics[width=0.24\linewidth,trim={3.6cm 1.3cm 3.2cm 1.3cm},clip]{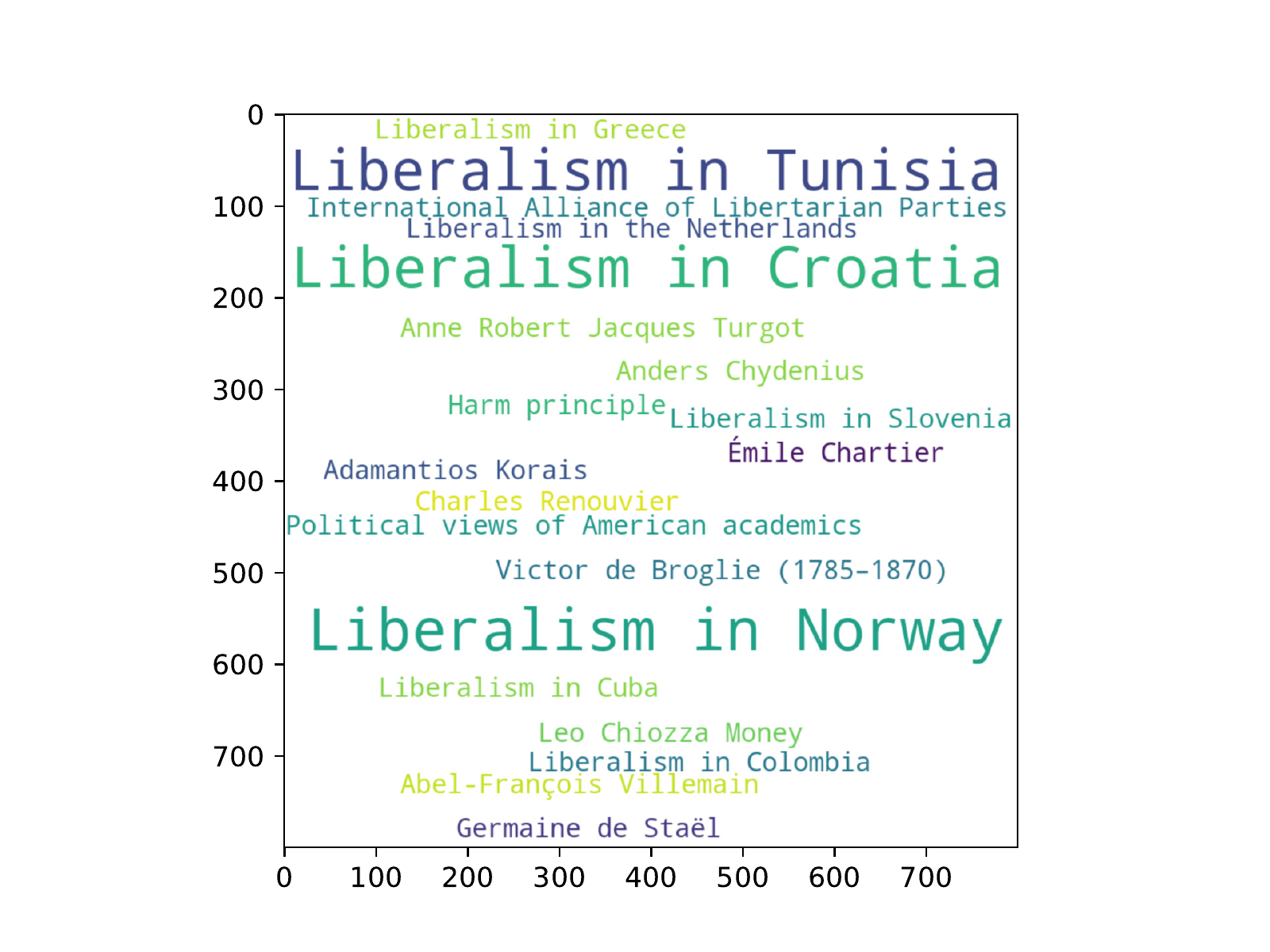}}
	\caption{Prominent members and covariates for two communities: panels (a) \& (b) regard a community that can be interpreted as \texttt{Asian Buddhism Philosophers} and panels (c) \& (d) correspond to a community that can be interpreted as \texttt{Political Liberalism Philosophers}.}
	\label{fig:philosophers}
\end{figure*}

\begin{table}
	\centering
	\caption{Community detection performance of \short{} on datasets with binary covariates. We report the mean and standard deviation in NMI scores. Cosine similarity kernel was used for spectral clustering when only node covariates are considered. The first block of methods uses traditional techniques that are not based on statistical models. Techniques in the next block use interpretable statistical models. The last block contains deep neural network-based approaches that are not explainable. Our method is highlighted in bold.}
	\label{table:community_detection_results}
	\begin{tabular}{lcc}
		\toprule 
		\textbf{Method} & \textbf{Cora} & \textbf{Citeseer}  \\
		\midrule
		\multicolumn{3}{l}{\textbf{\emph{Traditional methods}}} \\
		SC \cite{Luxburg:2007:ATutorialOnSpectralClustering} (only network) & $0.030$ & $0.019$ \\
		SC \cite{Luxburg:2007:ATutorialOnSpectralClustering} (only covariates) & $0.169$ & $0.202$ \\
		CASC \cite{BinkiewiczEtAl:2017:CovariateAssistedSpectralClustering} & $0.110$ & $0.182$ \\
		CODICIL \cite{RuanFuhryEtAl:2013:EfficientCommunityDetectionInLargeNetworksUsingContentAndLinks} & $0.368$ & $0.286$ \\
		CDE \cite{LiEtAl:2018:CommunityDetectionInAttributedGraphsAnEmbeddingsApproach} & $0.504$ & $0.299$ \\
		\midrule
		\multicolumn{3}{l}{\textbf{\emph{Explainable methods}}} \\
		CESNA \cite{YangEtAl:2013:CommunityDetectionInNetworksWithNodeAttributes} & $0.269$ & $0.022$ \\
		LLW \cite{EroshevaEtAl:2004:MixedMembershipModelsOfScientificPublications} & $0.359$ & $0.192$ \\
		PCL-PLSA \cite{YangEtAl:2009:CombiningLinkAndContentForCommunityDetection} & $0.390$ & $0.220$ \\
		PCL-DC \cite{YangEtAl:2009:CombiningLinkAndContentForCommunityDetection} & $0.512$ & $0.292$ \\
		\textbf{\short{} (Gibbs)} & $\mathbf{0.521 \pm 0.008}$ & $\mathbf{0.412 \pm 0.007}$ \\
		\textbf{\short{} (Exact)} & $\mathbf{0.511 \pm 0.01}$ & $\mathbf{0.401 \pm 0.008}$ \\ 
		\midrule
		\multicolumn{3}{l}{\textbf{\emph{Neural network based methods}}} \\
		DeepWalk \cite{PerozziEtAl:2014:DeepWalk} & $0.327$ & $0.088$ \\
		GAE \cite{KipfEtAl:2016:VariationalGraphAutoEncoders} & $0.429$ & $0.176$ \\
		VGAE \cite{KipfEtAl:2016:VariationalGraphAutoEncoders} & $0.436$ & $0.156$ \\
		ARGE \cite{PanEtAl:2018:AdversariallyRegularizedGraphAutoencoderForGraphEmbedding} & 0.449 & 0.359 \\
		ARVGE \cite{PanEtAl:2018:AdversariallyRegularizedGraphAutoencoderForGraphEmbedding} & 0.450 & 0.261 \\
		DAEGC \cite{WangEtAl:2019:AttributedGraphClustering} & 0.528 & 0.397 \\
		\bottomrule
	\end{tabular}
\end{table}

\begin{table}
	\centering
	\caption{Community detection performance of \short{} on the Sinanet network that has continuous covariates.}
	\label{table:community_detection_cont}
	\begin{tabular}{lccc}
		\toprule
		\textbf{Method} & CESNA \cite{YangEtAl:2013:CommunityDetectionInNetworksWithNodeAttributes} & Binary \short{} & Cont. \short{} \\
        \midrule
        \textbf{NMI} & $0.19$ & $0.23$ & $0.25 \pm 0.01$ \\
		\bottomrule
	\end{tabular}
\end{table}

\paragraph{Community detection} We experimented with real-world networks with known ground-truth communities. Table \ref{table:community_detection_results} compares the performance of \short{} with existing approaches on datasets that have binary covariates (Cora and Citeseer). Our model outperforms all existing methods that are explainable (second block in Table \ref{table:community_detection_results}) and all but one deep neural network-based method (third block in Table \ref{table:community_detection_results}) that are not explainable. We also experimented with the Sinanet network that has continuous covariates. Table \ref{table:community_detection_cont} compares the continuous variant of \short{} (Section \ref{section:continuous_rbsbm}) with the regular binary variant. To use the binary variant, we binarize the covariates by dividing the range of their values into ten equally sized bins. A continuous value is replaced with a one-hot encoded vector representing the bin to which that value belongs. Such binarization loses information, as is evident from Table \ref{table:community_detection_cont} where the continuous variant of \short{} outperforms the binary variant.

\begin{table}[t]
	\centering
	\caption{Link prediction performance of \short{}. We report mean and standard deviation in AUC scores. The first block of methods uses traditional approaches, and the last block of methods is based on neural networks. The performance of our model (highlighted in bold) is comparable with deep neural network-based approaches. However, \short{} comes with the additional advantage of explainability. A $\dagger$ denotes that the method does not use covariates.}
	\label{table:link_prediction_results}
    \resizebox{\linewidth}{!}{%
	\begin{tabular}{lccc}
		\toprule
		\textbf{Method} & \textbf{Cora} & \textbf{Citeseer}  & \textbf{PubMed} \\
		\midrule
		\multicolumn{4}{l}{\textbf{\emph{Explainable methods}}} \\
		CESNA \cite{YangEtAl:2013:CommunityDetectionInNetworksWithNodeAttributes} & $0.76 \pm 0.02$ & $0.65 \pm 0.03$ & $0.78 \pm 0.01$ \\
		SocioDim \cite{TangEtAl:2011:LeveragingSocialMediaNetworksForClassification} & $0.846 \pm 0.01$ & $0.805 \pm 0.01$ & $0.839 \pm 0.01$ \\
		\textbf{\short{}} & $\mathbf{0.884 \pm 0.006}$ & $\mathbf{0.876 \pm 0.01}$ & $\mathbf{0.861 \pm 0.02}$ \\
		\midrule
		\multicolumn{4}{l}{\textbf{\emph{Neural network based methods}}} \\
		DeepWalk \cite{PerozziEtAl:2014:DeepWalk} & $0.831 \pm 0.01$ & $0.85 \pm 0.02$ & $0.841 \pm 0.02$ \\
		$\dagger$GAE \cite{KipfEtAl:2016:VariationalGraphAutoEncoders}  & $0.843 \pm 0.02$ & $0.787 \pm 0.02$ & $0.874  \pm 0.01$ \\
		$\dagger$VGAE \cite{KipfEtAl:2016:VariationalGraphAutoEncoders}  & $0.84 \pm 0.02$ & $0.789 \pm 0.03$ & $0.875 \pm 0.01$ \\
		GAE \cite{KipfEtAl:2016:VariationalGraphAutoEncoders} & $0.91 \pm 0.02$ & $0.895 \pm 0.04$  & $0.965 \pm 0.01$\\
		VGAE \cite{KipfEtAl:2016:VariationalGraphAutoEncoders} & $0.914 \pm 0.01$ & $0.926 \pm 0.01$  & $0.947 \pm 0.02$ \\
		\bottomrule
	\end{tabular}}
\end{table}

\paragraph{Link prediction} \short{} is community focused but our experiments demonstrate that it also has good link-prediction performance. To perform link prediction, we select $20\%$ of the edges and an equal number of non-edges and mark them as unobserved. The remaining data is used for inferring the community memberships, block matrix, and other parameters of the model using the algorithm in Fig.~\ref{alg:inference_procedure}. The probability of missing links is obtained from the block matrix using the inferred community memberships of the nodes (the inference has already taken covariates into consideration). Table \ref{table:link_prediction_results} reports the AUC scores (AUC stands for Area Under Curve; higher values are better) for various methods. \short{} outperforms traditional approaches like CESNA \cite{YangEtAl:2013:CommunityDetectionInNetworksWithNodeAttributes} with a large margin. It even outperforms some of the deep neural network-based methods  and is only slightly worse than others while being more explainable. Note that PubMed has continuous covariates, so we use the continuous variant of \short{} for PubMed.

\begin{figure}
	\centering
    \subfloat[][$n=100$, $m=5$, $k=3$]{\includegraphics[width=0.24\textwidth,trim={0.5cm 0.8cm 0.5cm 1cm},clip]{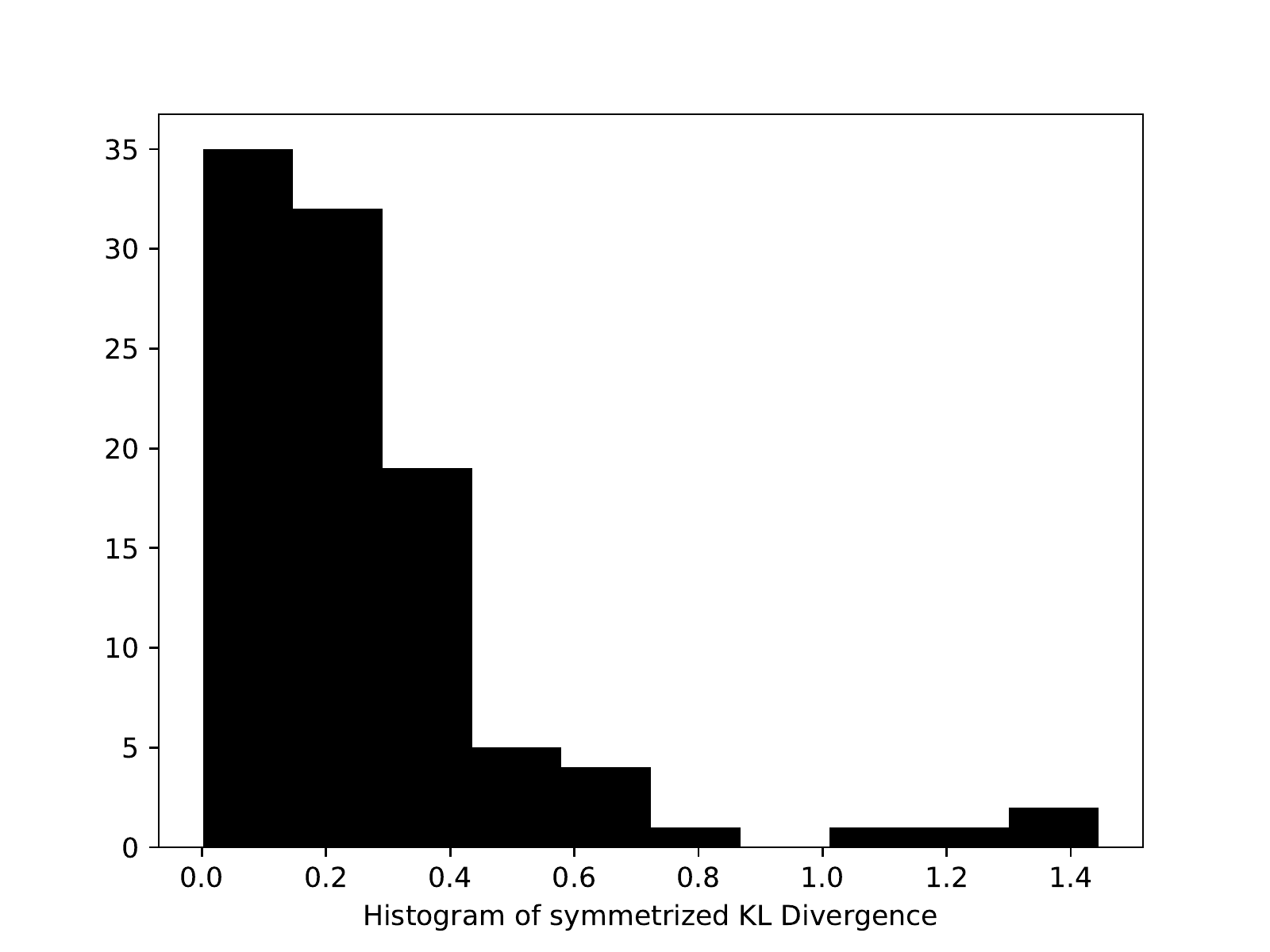}}%
    \subfloat[][$n=500$, $m=7$, $k=5$]{\includegraphics[width=0.24\textwidth,trim={0.5cm 0.8cm 0.5cm 1cm},clip]{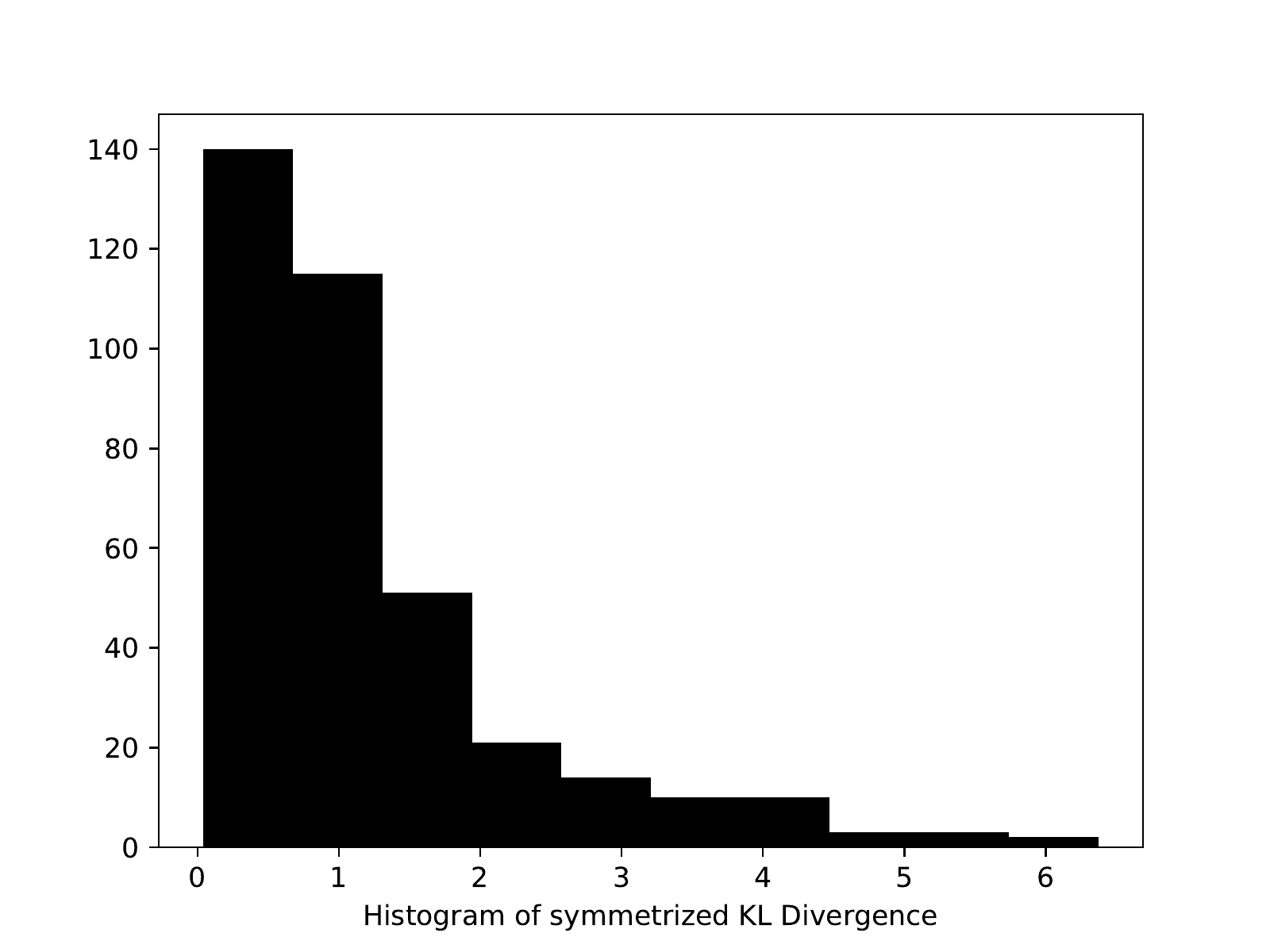}}	
	\caption{Histograms showing the Jensen-Shannon divergence between the sampled and recovered mixed-membership vectors $\bfZ$ for two configurations of SynthMM-$n$-$m$-$k$. The $x$-axis represents the divergence value and $y$-axis represents the number of nodes with that divergence value. Many nodes have very small divergence implying that the recovered vectors are very close to the sampled ones.}
	\label{fig:mmrsbm_synthetic_param}
\end{figure}

\begin{figure*}
	\centering
	\subfloat[][Sampled $\bfW$ (config. 1)]{\includegraphics[width=0.25\linewidth, trim={1cm 0.7cm 3.5cm 1cm}, clip]{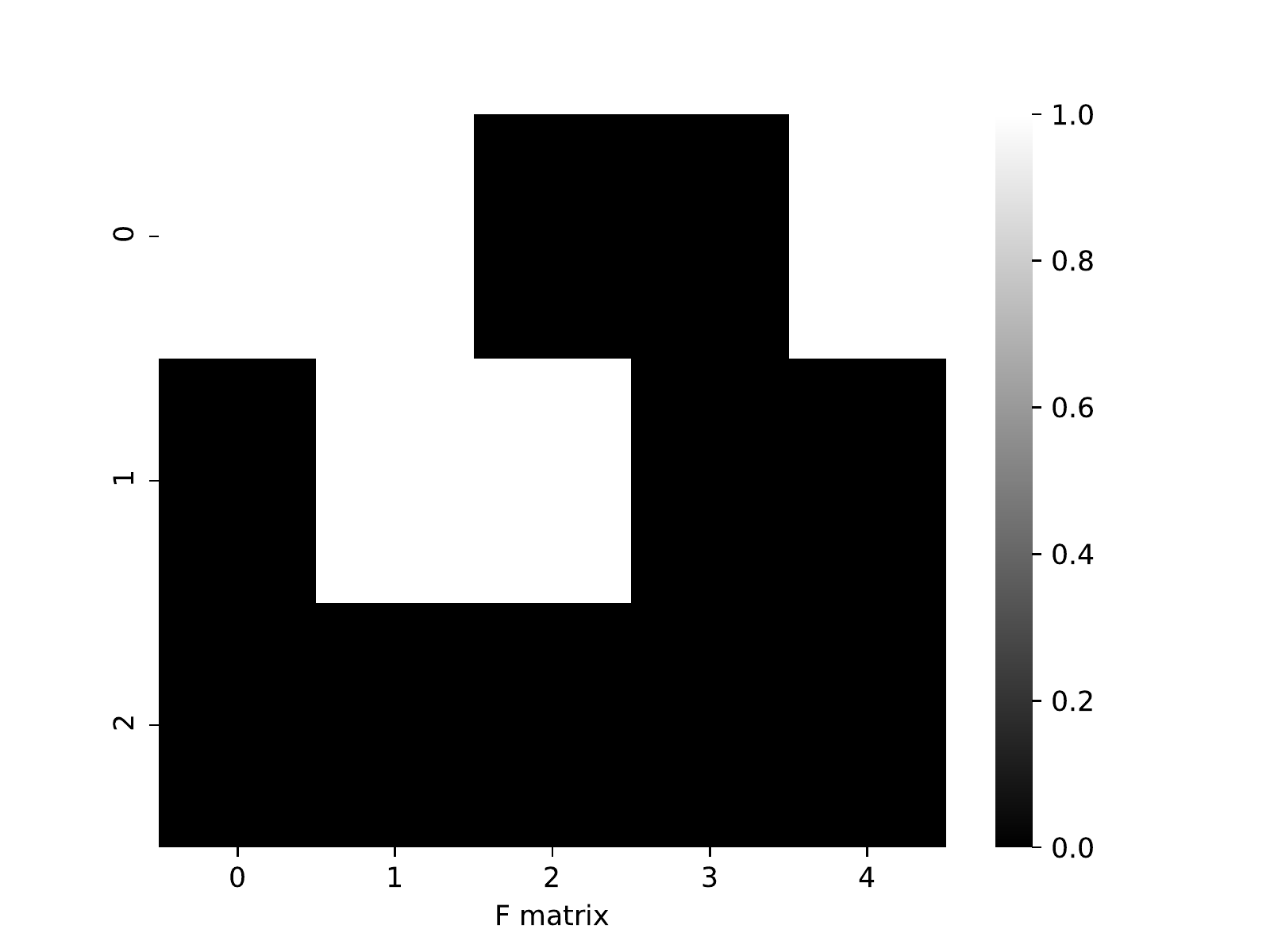}}%
	\subfloat[][Inferred $\bfW$ (config. 1)]{\includegraphics[width=0.25\linewidth, trim={1cm 0.7cm 3.5cm 1cm}, clip]{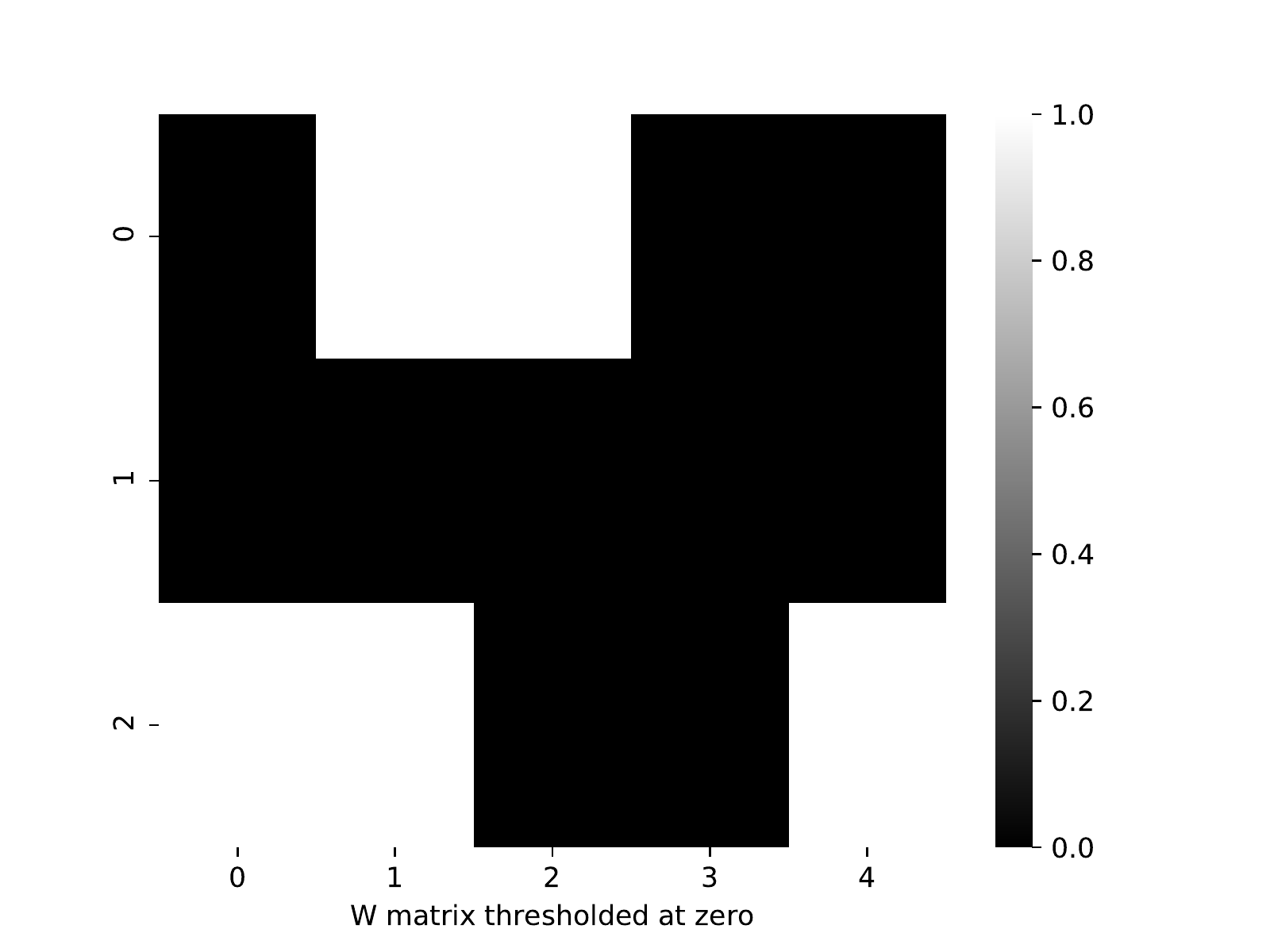}}%
	\subfloat[][Sampled $\bfW$ (config. 2)]{\includegraphics[width=0.25\linewidth, trim={1cm 0.7cm 3.5cm 1cm}, clip]{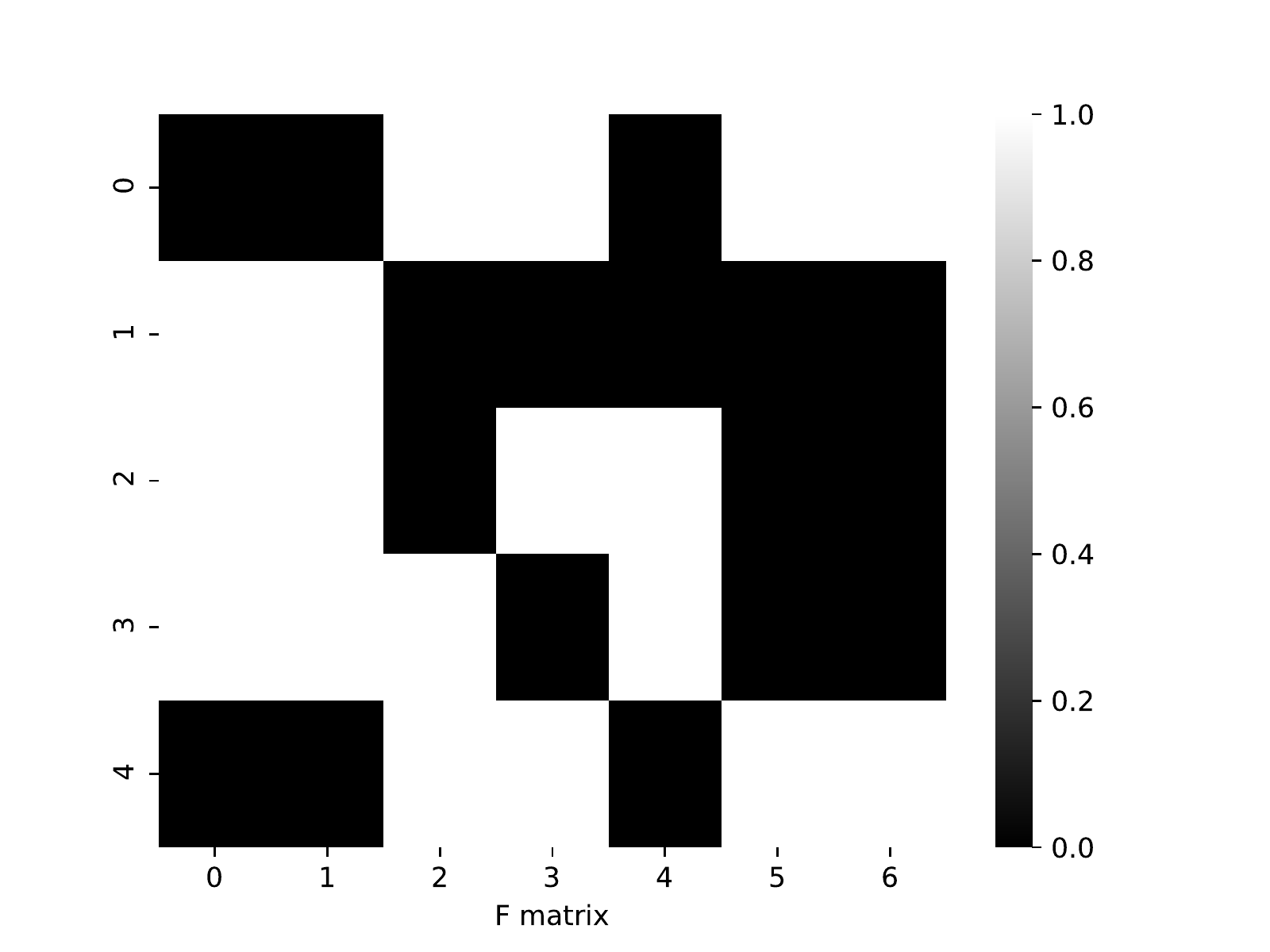}}%
	\subfloat[][Inferred $\bfW$ (config. 2)]{\includegraphics[width=0.25\linewidth, trim={1cm 0.7cm 3.5cm 1cm}, clip]{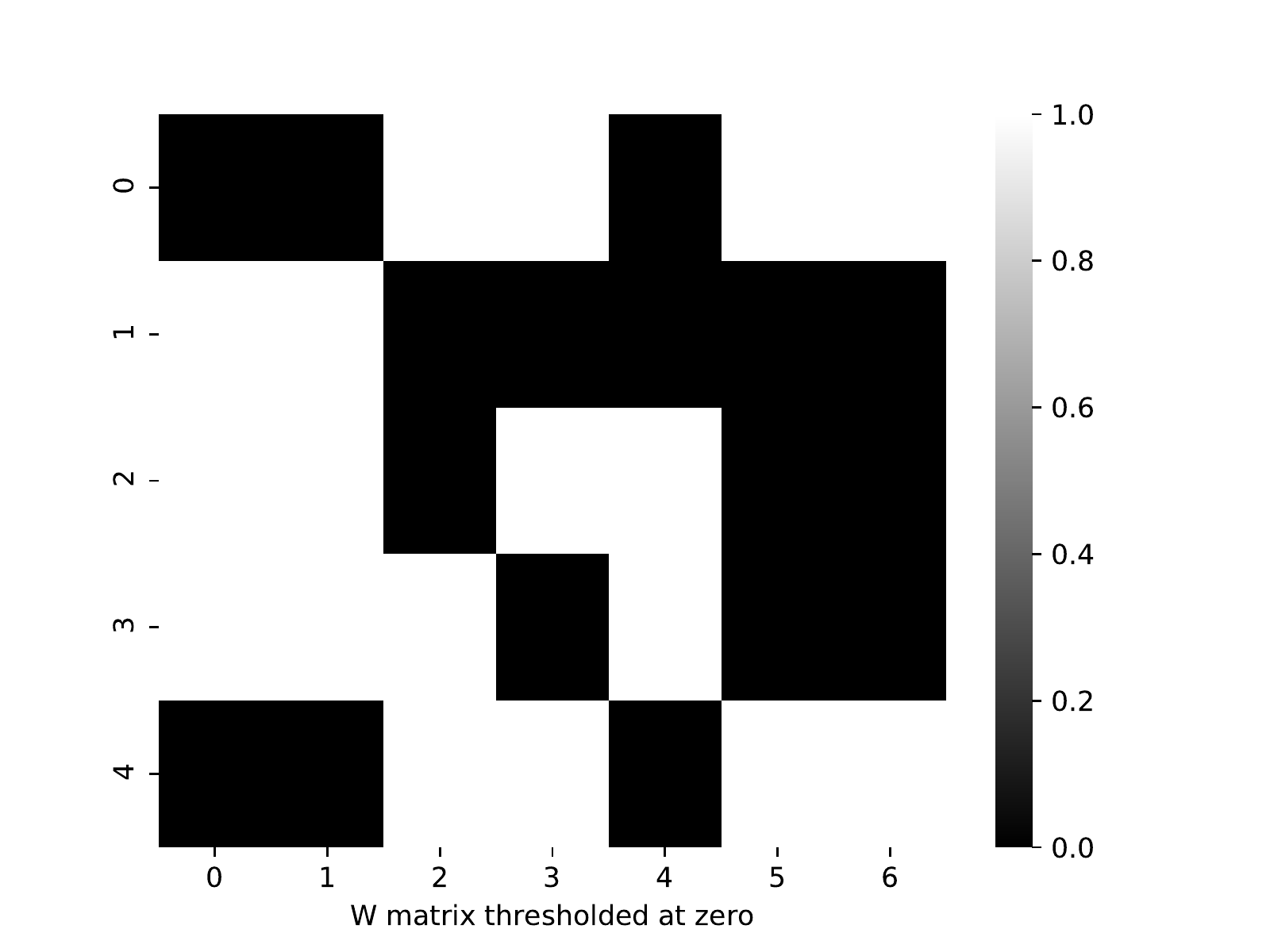}}
	\caption{Comparison between the sampled and inferred RBM weight matrix $\bfW$ for two configurations of SynthMM-$n$-$m$-$k$. Config. 1: $n = 100$, $m = 5$, and $k = 3$. Config. 2: $n = 500$, $m = 7$, and $k = 5$. Entries have been thresholded at $0.5$ for better clarity. The sampled and inferred matrices are same upto a permutation of rows in both cases.}
	\label{fig:mmrsbm_synthetic_W}
\end{figure*}

\paragraph{Explainability} In this experiment, we take the Philosophers network where ground-truth communities are unknown, hence metrics like NMI can no longer be computed. Therefore, an approach that gives interpretable insights underlying the discovered communities is particularly useful. Inference in \short{} identifies the salient covariates in each community using the $\bfW$ matrix from RBM (see Section \ref{section:rbsbm_description}). This provides an explanation for what holds the communities together from the perspective of the model. Due to space constraints, we only present two such communities in Figure~\ref{fig:philosophers} (out of $k=20$ discovered communities). Figure~\ref{fig:philosophers} shows the members and covariates of two communities that can be interpreted as \texttt{Asian Buddhism Philosophers} and \texttt{Political Liberalism Philosophers}. Let $\ell$ denote the index of the community under consideration. The size of the text in the word-cloud is proportional to the value of $z_{i\ell}$ for community memberships and $W_{j\ell}$ for salient covariates. \short{} was able to discover meaningful communities while at the same time highlighting the importance of various covariates in them. While approaches like \cite{YangEtAl:2013:CommunityDetectionInNetworksWithNodeAttributes} can also provide such insights, Table \ref{table:community_detection_results} shows that they have poor quantitative performance in practice. \short{} stands out because it performs comparably to modern deep neural network-based methods while also offering explainability.


\begin{figure}
	\centering
	\includegraphics[width=\linewidth,trim={2cm 2cm 2cm 2.5cm},clip]{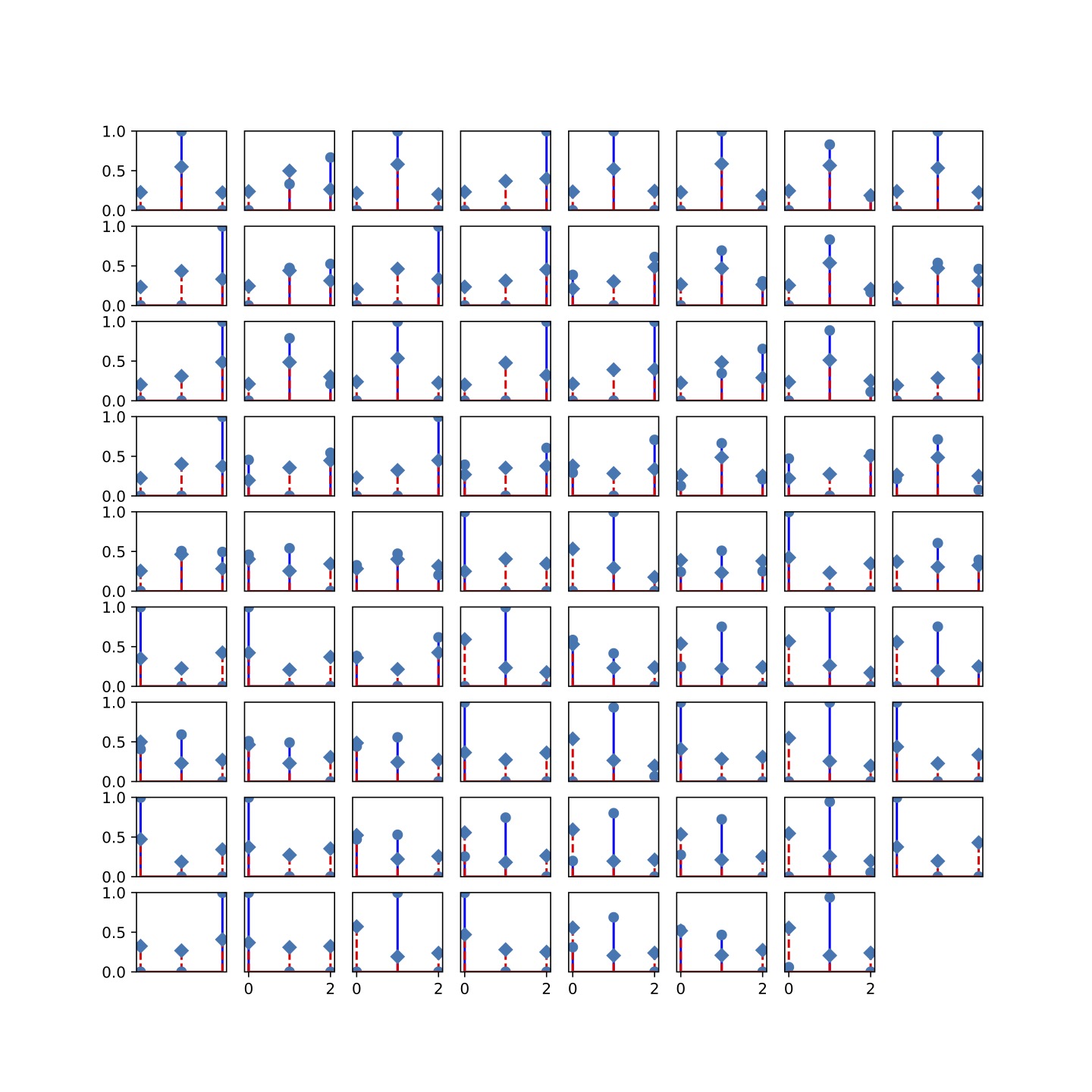}
	\caption{Comparison between the mixed-membership vectors inferred by \shortmm{} and mixed-membership SBM for the Lazega Lawyers dataset. Each cell corresponds to a node. The height of the bars represent the value of the corresponding entries of $\bfz_i$ (we use $k = 3$) inferred by \shortmm{} (circle) and mixed-membership SBM (diamond).}
	\label{fig:lazega_mmvec_plot}
\end{figure}

\begin{figure}
	\centering
	\subfloat[][\shortmm{}]{\includegraphics[width=0.5\linewidth]{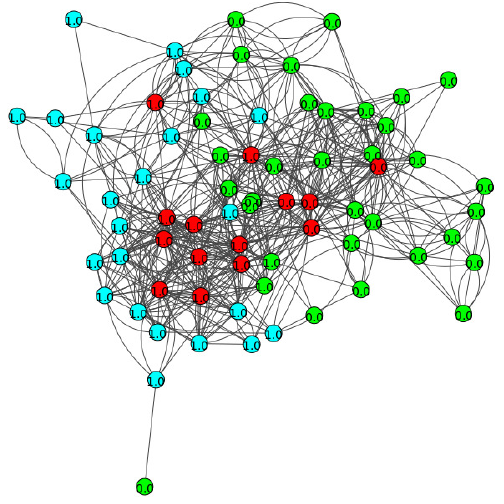}}
	\subfloat[][Mixed-membership SBM \cite{Airoldi:2008:MixedMembershipStochasticBlockmodels}]{\includegraphics[width=0.5\linewidth]{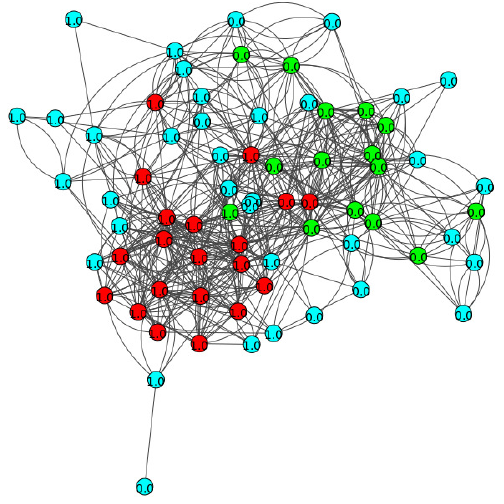}}
	\caption{Comparison between communities discovered by \shortmm{} and mixed-membership SBM. Nodes are colored based on the most likely community under their mixed-membership vector $\bfz_i$ ($k = 3$). \shortmm{} discovers more meaningful communities that are explained by the attributes. See Section \ref{section:experiments_rbmmsbm} for details.}
	\label{fig:lazega_network}
\end{figure}

\subsection{Results for \shortmm{}}
\label{section:experiments_rbmmsbm}

All experiments up to this point involved networks where nodes are assumed to belong to exactly one community. In this section, we use networks with mixed-community memberships to evaluate the performance of \shortmm{}. As before, we begin by establishing the correctness of the inference procedure and then show that \shortmm{} can discover meaningful communities by leveraging both covariates and links and provide explanations for these communities.

\paragraph{Correctness} We sampled synthetic networks SynthMM-$n$-$m$-$k$ from \shortmm{} using the following two configurations: \textbf{(i)} $n = 100$, $m = 5$, and $k = 3$, and \textbf{(ii)} $n = 500$, $m = 7$, and $k = 5$. Then, as before, we used the inference procedure to recover the mixed-membership vectors $\bfZ$, block matrix $\bfB$, and RBM parameters $\bfW$, $\bfu$, and $\bfv$ from the observed data ($\bfA$ and $\bfY$). To evaluate the discovered communities, we computed the symmetrized KL-divergence (also known as Jensen-Shannon divergence) between the ground-truth and discovered mixed-membership vectors $\bfZ$. Figure \ref{fig:mmrsbm_synthetic_param} shows this metric in a histogram for both the configurations listed above. The majority of the nodes have a very low divergence value, and hence their mixed-membership vectors are close to the ground-truth vectors. We similarly observed that the mean absolute difference between the recovered and sampled block matrix $\bfB$ is about $0.065$ in both cases. Finally, Figure \ref{fig:mmrsbm_synthetic_W} compares the recovered matrix $\bfW$ to the original matrix. We have thresholded the entries of the matrices at $0.5$ for a cleaner presentation. As expected, the recovered $\bfW$ is the same as the sampled $\bfW$ up to a permutation of rows for both configurations.

\paragraph{Community detection and explainability} Our last experiment uses the Lazega Lawyers dataset. We experimented with different values of $k$ between $2$ and $8$, and selected $k = 3$ as it maximizes \textrm{ELBO}. Figure \ref{fig:lazega_mmvec_plot} compares the mixed-membership vectors discovered by \shortmm{} and mixed-membership SBM for $k = 3$. Figure \ref{fig:lazega_network} shows the most likely community for each node, as inferred by \shortmm{} and mixed-membership SBM. One of the covariates indicates whether the node is a `partner' (labeled $0$) or an `associate' (labeled $1$) (Appendix \ref{appendix:experiments}). In communities discovered by \shortmm{}, most members of the blue community are partners, and most members of the green community are associates. The red community has high-degree nodes (minimum degree = $15$) and contains both partners and associates that act as bridges between the green and blue communities. Such a clear interpretation is missing in the communities discovered by the mixed-membership SBM.


\section{Conclusion}
\label{section:conclusion}
This paper addresses the problem of explainable inference and modeling networks with covariates. The key idea is to enhance the stochastic block model and variants with a restricted Boltzmann machine to capture the relation between node covariates and possible communities in the network. The appeal of this approach is that it yields models that offer explanations for the discovered communities without compromising on the real-world quantitative performance on tasks like community detection and link prediction. We developed efficient inference procedures for the proposed models. For \short{}, the proposed iterative procedure is rather scalable, as each iteration runs in time linear in the number of nodes and links in the network. For the more complicated \shortmm{} model, the complexity of each iteration is higher, but it is still possible to deal with moderately sized networks. 

A challenging but very useful direction for future research would be to quantify (in a statistically sound way) the uncertainty associated with the learned parameter models. Furthermore, owing to the generative nature of the proposed models, it is possible to create synthetic network data that can be used as a benchmark to test different inference methods for community detection and/or link prediction. We believe that our proposed models serve as a stepping-stone for the generalization of SBMs to networks with node and link covariates. For instance, one can consider other variants of SBMs like the Degree-Corrected SBM \cite{KarrerNewman:2011:StochasticBlockmodelsAndCommunityStructureInNetworks} to better model much more realistic heterogeneous communities. Likewise, models similar in spirit to RBM can be developed to incorporate link covariates as well, and one expects that similar variational EM approaches can be developed and would lead to efficient and scalable inference procedures for those extensions.


\appendices

\section{\short{}: Additional Details}

This appendix presents additional technical details related to \short{}.


\subsection{Normalization Constant for RBM}
\label{appendix:normalization_constant_rbm}

In a usual RBM, both the observed and hidden units are binary. Computing the normalization constant requires adding an exponential number of terms\footnote{Exponential in the number of observed and hidden units in the RBM} in such a case, making it intractable \cite{FischerEtAl:2012:AnIntroductionToRestrictedBoltzmannMachines}. We use a modified variant of the RBM where the units corresponding to the community membership are one-hot encoded. Below, we show that this simplification allows us to efficiently compute the normalization constant $\Psi(\bfW, \bfu, \bfv)$ in \eqref{eq:partition_function_computation_main} in $O(km)$ steps.

Recall that the RBM models a joint distribution over random vectors $\bfy \in \{0, 1\}^m$ and $\bfz \in \{0, 1\}^k$ such that $\mathbf{1}^\intercal \bfz = 1$. We begin by computing the marginal distribution over $\bfz$. Let $\bfz$ be such that $z_\ell = 1$ for an arbitrarily chosen $\ell \in \{1, 2, \dots, k\}$.
{
\small
\begin{align*}
	\rmP_{\bm{\theta}}(\bfz) &= \sum_{\bfy \in \{0, 1\}^m} \rmP_{\bm{\theta}}(\bfy, \bfz) \\
	&= \frac{1}{\Psi(\bfW, \bfu, \bfv)} \sum_{\bfy \in \{0, 1\}^m} \exp(\sum_{j = 1}^m (W_{j\ell} + u_j)y_j + v_\ell) \\
	&= \frac{\exp(v_\ell)}{\Psi(\bfW, \bfv, \bfu)} \sum_{\bfy \in \{0, 1\}^m} \prod_{j=1}^m \exp((W_{j\ell} + u_j) y_j) \\
	&= \frac{\exp(v_\ell)}{\Psi(\bfW, \bfv, \bfu)} \prod_{j=1}^m (1 + \exp(W_{j\ell} + u_j)).
\end{align*}
}
Because $\sum_{\ell=1}^k \rmP_{\bm{\theta}}(z_\ell = 1) = 1$, we get,
\begin{equation*}
    \Psi(\bfW, \bfv, \bfu) = \sum_{\ell=1}^k \exp(v_\ell) \prod_{j=1}^m (1 + \exp(W_{j\ell} + u_j)).
\end{equation*}


\subsection{Exact Computation of RBM Gradients}
\label{appendix:exact_gradient_computation}

To update the RBM parameters, one needs to compute the expectations in \eqref{eq:gradients}. These expectations are computed using the Monte-Carlo method in a usual RBM by drawing samples via Gibbs sampling (see \eqref{eq:conditional_distributions}). However, we can also get closed-form expressions for these expectations in our case, as the normalization constant $\Psi(\bfW, \bfu, \bfv)$ can be computed efficiently. First, note that
\begin{align*}
    \rmE_{\rmP_{\mathrm{RBM}}}[y_j z_\ell] &= \rmP_{\bm{\theta}}(y_j = 1, z_\ell = 1) \\
    &= \rmP_{\bm{\theta}}(y_j = 1 \vert z_\ell = 1) \rmP_{\bm{\theta}}(z_\ell = 1), \\
    \rmE_{\rmP_{\mathrm{RBM}}}[z_\ell] &= \rmP_{\bm{\theta}}(z_\ell = 1), \\
    \rmE_{\rmP_{\mathrm{RBM}}}[y_j] &= \rmP_{\bm{\theta}}(y_j=1).
\end{align*}
The expressions for $\rmP_{\bm{\theta}}(y_j = 1 \vert z_\ell = 1)$ and $\rmP_{\bm{\theta}}(z_\ell = 1)$ are given in \eqref{eq:conditional_distributions} in Appendix \ref{appendix:normalization_constant_rbm}, respectively. $\rmP_{\bm{\theta}}(y_j=1 )$ can also be computed easily using $\rmP_{\bm{\theta}}(y_j = 1 \vert z_\ell = 1)$ and $\rmP_{\bm{\theta}}(z_\ell = 1)$ as follows:
\begin{equation*}
    \rmP_{\bm{\theta}}(y_j = 1) = \sum_{\ell=1}^k \rmP_{\bm{\theta}}(y_j=1 \vert z_\ell=1) \rmP_{\bm{\theta}}(z_\ell = 1).
\end{equation*}
Computing $\Psi(\bfW, \bfu, \bfv)$ in \eqref{eq:partition_function_computation_main} requires a product over $m$ terms. Thus, the exact computation of gradients is likely to encounter numerical stability issues, as $\Psi(\bfW, \bfu, \bfv)$ is needed to compute $\rmP_{\bm{\theta}}(z_\ell = 1)$ in the expectation terms above.


\section{\shortmm{}: Additional Details}

This appendix presents additional technical details related to \shortmm{}.


\subsection{Sampling from the modified RBM}
\label{appendix:sampling_from_modified_rbm_mmsbm}

In this section, we derive the conditional distributions in \eqref{eq:conditional_distributions_rbmmsbm} that are used for computing the gradients in \eqref{eq:gradients} via Gibbs sampling. For every vector $\bfy \in \{0, 1\}^m$, let $\bfy_{-j} \in \{0, 1\}^{m - 1}$ denote a vector that has all elements of $\bfy$ except the $j^{th}$ element. Now,
\begin{align*}
    \rmP_{\bm{\theta}}(y_j = 1 \vert \bfz) &= \rmP_{\bm{\theta}}(y_j = 1 \vert \bfz, \bfy_{-j}) \\
    &= \frac{\rmP_{\bm{\theta}}(y_j = 1, \bfz, \bfy_{-j})}{\sum_{y \in \{0, 1\}} \rmP_{\bm{\theta}}(y_j = y, \bfz, \bfy_{-j})} \\
    &=  \frac{\exp(\sum_{\ell = 1}^k W_{j\ell} z_\ell + u_j)}{1 + \exp(\sum_{\ell = 1}^k W_{j\ell} z_\ell + u_j)} \\
    &= \sigma(\sum_{\ell = 1}^k W_{j\ell} z_\ell + u_j),
\end{align*} 
where $\sigma(x) = \frac{1}{1 + \exp(-x)}$ is the logistic sigmoid function. 

Sampling $\bfz$ given $\bfy$ is more tricky as the elements of $\bfz$ must add up to one. Notice that $z_\ell$ is uniquely determined if all elements of $\bfz_{-\ell}$ are held constant. Thus, we sample two elements of $\bfz$ at a time. Without loss of generality, let us sample $z_\ell$ and $z_k$ for an arbitrary $\ell \in \{1, 2, \dots, k - 1\}$. Let $\bar{\bfz}_{-\ell}$ denote all entries of vector $\bfz$ except the $\ell^{th}$ and $k^{th}$ entries. Define $s_\ell = \mathbf{1}^\intercal \bar{\bfz}_{-\ell}$ to be the sum of all entries in $\bar{\bfz}_{-\ell}$ and let $\beta_{\ell} = v_\ell - v_k + \sum_{j = 1}^m (W_{j\ell} - W_{jk})y_j$. When $z_{\ell} = z$ for some $z \in [0, 1 - s_{\ell}]$, $z_{k} = 1 - z - s_{\ell}$. Thus, by sampling $z$, we can update both $z_{\ell}$ and $z_{k}$. To sample $z$, we use $\rmP_{\bm{\theta}}(z_\ell = z, z_k = 1 - z - s_{\ell} \vert  \bfy, \bar{\bfz}_{-\ell})$.
{
\small
\begin{align*}
    \rmP_{\bm{\theta}}(z_\ell = z, &z_k = 1 - z - s_{\ell} \vert  \bfy, \bar{\bfz}_{-\ell}) \\
    &= \frac{\rmP_{\bm{\theta}}(z_\ell = z, z_k = 1 - z - s_{\ell},  \bfy, \bar{\bfz}_{-\ell})}{\int_0^{1-s_\ell} \rmP_{\bm{\theta}}(z_\ell = z{'}, z_k = 1 - z{'} - s_{\ell},  \bfy, \bar{\bfz}_{-\ell}) \mathrm{d}z{'}} \\
    &= \frac{\exp(z \beta_\ell)}{\int_0^{1-s_\ell} \exp(z{'} \beta_\ell) \mathrm{d}z{'}} \\
    &= \frac{\beta_\ell \exp(z \beta_\ell)}{\exp((1-s_\ell)\beta_\ell) - 1}.
\end{align*}
}%
A vector $\bfz$ can be sampled from the distribution above by iterating over the indices $\ell = 1, 2, \dots, k - 1$.


\subsection{Derivation of \textrm{ELBO}}
\label{appendix:rbmmsbm_elbo_derivation}

Section \ref{section:inference_in_rbmmsbm} specifies assumptions on the parametric form of various factors in the approximate posterior $\rmQ$. Under those assumptions, \textrm{ELBO} can be computed as
{
\small
\begin{align*}
    \calL_{\rmQ}(\bm{\theta}) &= \rmE_{\rmQ}[\ln \rmP_{\bm{\theta}}(\bfA, \bfY, \bfZ, \bfPsi, \bfB)] - \ln \rmQ(\bfZ, \bfPsi, \bfB)] \\
    &= \rmE_{\rmQ}[\ln \rmP_{\bm{\theta}}(\bfB)] + \rmE_{\rmQ}[\ln \rmP_{\bm{\theta}}(\bfY, \bfZ)] + \rmE_{\rmQ}[\ln \rmP_{\bm{\theta}}(\bfPsi | \bfZ)]\\
    &\;\;\; + \rmE_{\rmQ}[\ln \rmP_{\bm{\theta}}(\bfA \vert \bfPsi, \bfB)] - \rmE_{\rmQ}[\ln \rmQ(\bfZ, \bfPsi, \bfB)].
\end{align*}
}%
Next, we compute each term in this expectation separately.
{
\small
\begin{align*}
    \rmE_{\rmQ}[\ln \rmP_{\bm{\theta}}&(\bfB)] = \sum_{i, j = 1}^k \rmE_{q_{ij}}[\ln \rmP_{\bm{\theta}}(B_{ij})] \\
    &= \sum_{i, j = 1}^k \rmE_{q_{ij}}[(\alpha_{ij} - 1) \ln B_{ij} + (\beta_{ij} - 1)\ln (1 - B_{ij})] \\
    &\;\;\;-\sum_{i, j = 1}^k B(\alpha_{ij}, \beta_{ij}) \\
    &= \sum_{i, j = 1}^k (\alpha_{ij} - 1)(\Psi(\bar{\alpha_{ij}}) - \Psi(\bar{\alpha}_{ij} + \bar{\beta}_{ij})) \\
    &\;\;\;+ \sum_{i, j = 1}^k (\beta_{ij} - 1)(\Psi(\bar{\beta}_{ij}) - \Psi(\bar{\alpha}_{ij} + \bar{\beta}_{ij})) + \mathrm{const}_4.
\end{align*}
}%
Here, recall that $\bar{\alpha}_{ij}$ and $\bar{\beta}_{ij}$ are parameters of $q_{ij}$, which is assumed to be a Beta distribution. The term $\mathrm{const}_4$ includes all terms that do not depend on the parameters of $\rmQ$. As before, $B(.)$ denotes the Beta function and $\Psi(.)$ denotes the digamma function. For the next term, recall that $q_{i}$ is assumed to be a Dirichlet distribution with parameters $[\mu_{i1}, \mu_{i2}, \dots, \mu_{ik}]$.
{
\small
\begin{align*}
    \rmE_{\rmQ}[\ln &\rmP_{\bm{\theta}}(\bfY, \bfZ)] = \sum_{i=1}^n \rmE_{q_i}[\ln \rmP_{\bm{\theta}}(\bfy_i, \bfz_i)] \\
    &= \sum_{i=1}^n \sum_{j=1}^m \sum_{\ell=1}^k Y_{ij} W_{j\ell}\frac{\mu_{i\ell}}{\sum_{\ell{'} = 1}^k \mu_{i\ell{'}}}
    + \sum_{i=1}^n \sum_{j=1}^m u_j Y_{ij} \\
    &\;\;\;+ \sum_{i=1}^n \sum_{\ell=1}^k v_\ell\frac{\mu_{i\ell}}{\sum_{\ell{'} = 1}^k \mu_{i\ell{'}}} - n \ln \Omega(\bfW, \bfu, \bfv).
\end{align*}
}%
We have used the fact that $E[X_i] = \frac{\mu_i}{\sum_{j = 1}^k \mu_j}$ if $X \sim \mathrm{Dirichlet}(\mu_1, \dots, \mu_k)$. Consider the next term,
{
\small
\begin{align*}
    \rmE_{\rmQ}[\ln \rmP_{\bm{\theta}}(\bfPsi \vert \bfZ)] &= \sum_{i, j = 1}^n \rmE_{\rmQ}[\ln \rmP_{\bm{\theta}}(\bfpsi_{ij}^{(i)} \vert \bfz_i)] + \rmE_{\rmQ}[\ln \rmP_{\bm{\theta}}(\bfpsi_{ij}^{(j)} \vert \bfz_j)] \\
    &= \sum_{i, j = 1}^n \sum_{\ell = 1}^k \Big[ \rmE_{q_{ij}^{(i)}}[\bfpsi_{ij}^{(i)}(\ell)] \rmE_{q_i}[\ln Z_{i\ell}] \\
    &\;\;\;+ \rmE_{q_{ij}^{(j)}}[\bfpsi_{ij}^{(j)}(\ell)] \rmE_{q_j}[\ln Z_{j\ell}] \Big] \\
    &= \sum_{i, j = 1}^n \sum_{\ell = 1}^k \Big[\alpha_{ij}^{(i)}(\ell) \Big(\Psi(\mu_{i\ell}) - \Psi(\sum_{\ell{'} = 1}^k \mu_{i\ell{'}}) \Big) \\
    &\;\;\;+ \alpha_{ij}^{(j)}(\ell) \Big(\Psi(\mu_{j\ell}) - \Psi(\sum_{\ell{'} = 1}^k \mu_{j\ell{'}}) \Big) \Big].
\end{align*}
}%
The last equality follows as $\rmE[\ln X_i] = \Psi(\mu_i) - \Psi(\sum_{j = 1}^k \mu_j)$ if $X \sim \mathrm{Dirichlet}(\mu_1, \dots, \mu_k)$. Recall that $\alpha_{ij}^{(i)}(\ell)$ is the probability with which $\bfpsi_{ij}^{(i)}(\ell)$ is one under the multinomial distribution $q_{ij}^{(i)}$. Next,
{
\small
\begin{align*}
    &\rmE_{\rmQ}[\ln \rmP_{\bm{\theta}}(\bfA \vert \bfPsi, \bfB] = \sum_{i \neq j} \rmE_{\rmQ}[\ln \rmP_{\bm{\theta}}(A_{ij} \vert \bfpsi_{ij}^{(i)}, \bfpsi_{ij}^{(j)}, \bfB)] \\
    &= \sum_{i \neq j} \sum_{\ell_1, \ell_2 =1}^k \Big[A_{ij} \rmE_{\rmQ}[\bfpsi^{(i)}_{ij}(\ell_1)] \rmE_{\rmQ}[\bfpsi^{(j)}_{ij}(\ell_2)] \rmE_{\rmQ}[\ln B_{\ell_1\ell_2}] \\
    &\;\;\;+ (1 - A_{ij}) \rmE_{\rmQ}[\bfpsi^{(i)}_{ij}(\ell_1)] \rmE_{\rmQ}[\bfpsi^{(j)}_{ij}(\ell_2)] \rmE_{\rmQ}[\ln (1 - B_{\ell_1\ell_2})] \Big] \\    
    &= \sum_{i \neq j} \sum_{\ell_1, \ell_2 =1}^k \Big[A_{ij} \alpha^{(i)}_{ij}(\ell_1) \alpha^{(j)}_{ij}(\ell_2) \big(\Psi(\bar{\alpha}_{\ell_1\ell_2}) - \Psi(\bar{\alpha}_{\ell_1 \ell_2} + \bar{\beta}_{\ell_1\ell_2})\big) \\
    &\;\;\;+ (1 - A_{ij}) \alpha^{(i)}_{ij}(\ell_1) \alpha^{(j)}_{ij}(\ell_2) \big(\Psi(\bar{\beta}_{\ell_1\ell_2}) - \Psi(\bar{\alpha}_{\ell_1 \ell_2} + \bar{\beta}_{\ell_1\ell_2})\big) \Big].
\end{align*}
}%
Finally, the last term simply computes the entropy of $\rmQ$. Using standard results for the entropy of beta, multinomial, and Dirichlet distributions, we get
{
\small
\begin{align*}
    &-\rmE_{\rmQ}[\ln \rmQ(\bfPsi, \bfZ, \bfB)] = \sum_{i, j = 1}^n \sum_{\ell = 1}^k \Big[\alpha_{ij}^{(i)}(\ell) \ln \alpha_{ij}^{(i)}(\ell) \\
    &\;\;\;+ \alpha_{ij}^{(j)}(\ell) \ln \alpha_{ij}^{(j)}(\ell)\Big] + \sum_{i=1}^n \Big[ \ln \bfB_F(\bfmu_i) \\
    &\;\;\;+ (\sum_{\ell = 1}^k \mu_{i\ell} - k) \Psi(\sum_{\ell = 1}^k \mu_{i\ell}) - \sum_{\ell = 1}^k (\mu_{i\ell} - 1)\Psi(\mu_{i\ell}) \Big] \\
    &\;\;\;+ \sum_{\ell_1, \ell_2 = 1}^k \Big[ \ln B(\bar{\alpha}_{\ell_1\ell_2}, \bar{\beta}_{\ell_1\ell_2}) - (\bar{\alpha}_{\ell_1\ell_2} - 1)\Psi(\bar{\alpha}_{\ell_1\ell_2}) \\
    &\;\;\;- (\bar{\beta}_{\ell_1\ell_2} - 1)\Psi(\bar{\beta}_{\ell_1\ell_2}) + (\bar{\alpha}_{\ell_1\ell_2} + \bar{\beta}_{\ell_1\ell_2} - 2)\Psi(\bar{\alpha}_{\ell_1\ell_2} + \bar{\beta}_{\ell_1\ell_2}) \Big].
\end{align*}
}%
Here, for $\bfx = [x_1, \dots, x_k]$, $\bfB_F(\bfx) = \frac{\prod_{i=1}^k \Gamma(x_i)}{\Gamma(\sum_{i = 1}^k x_i)}$ is the multivariate Beta function and $\Gamma(\cdot)$ is the gamma function. 

Together, these terms provide the expression for \textrm{ELBO}.


\section{Lazega Lawyers: Additional Details}
\label{appendix:experiments}

This section provides additional details for preprocessing the covariates in the Lazega Layers dataset to get binary covariates. Categorical covariates like gender (woman or man), status (partner or associate), office (Boston, Hartford, or Providence), practice (litigation or corporate), and law school (Harvard, Yale, and UCon) were converted to one-hot encoded vectors. The continuous covariate age was first converted to a categorical variable by assigning each value to one of the following bins: $[20, 30]$, $[31, 40]$, $[41, 50]$, $[51, 60]$, and $[61, 70]$. Then, this categorical data was represented using one-hot encoded vectors. Similarly, the continuous covariate `years with the firm' was also represented using one-hot encoded vectors by dividing the range of values ($0$--$35$) into equal-sized bins of size $5$. The resulting one-hot encoded vectors were concatenated to get a single binary covariate vector of size $24$ for each node. 


\bibliographystyle{IEEEtran}
\bibliography{biblio}


\begin{IEEEbiography}
    [{\includegraphics[width=1in,height=1.25in,clip]{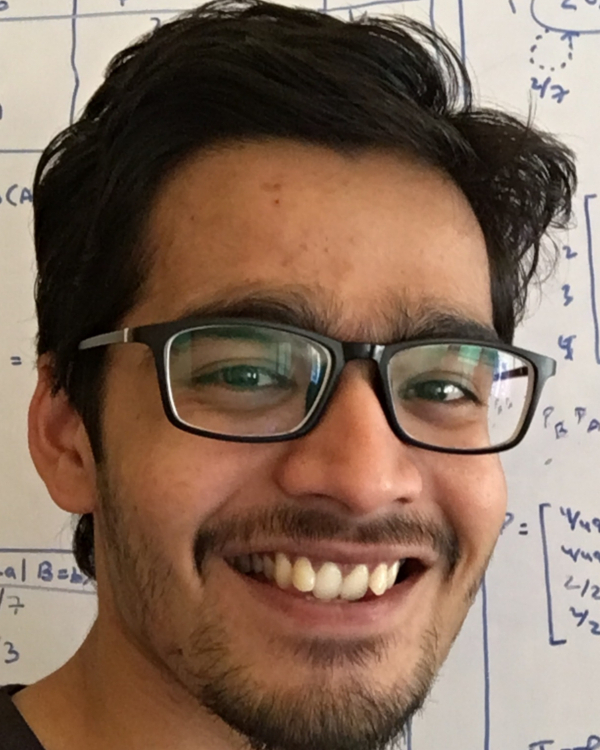}}]
    {Shubham Gupta} is a Ph.D. candidate in the Department of Computer Science and Automation at the Indian Institute of Science Bangalore. His primary research interests are in the field of multi-agent reinforcement learning, statistical analysis of dynamic networks, active learning for natural language processing, and one-shot learning in the visual domain. He is a recipient of the Wipro Ph.D. Fellowship (2018-2021).
\end{IEEEbiography}
    
\begin{IEEEbiography}
    [{\includegraphics[width=1in,height=1.25in,clip]{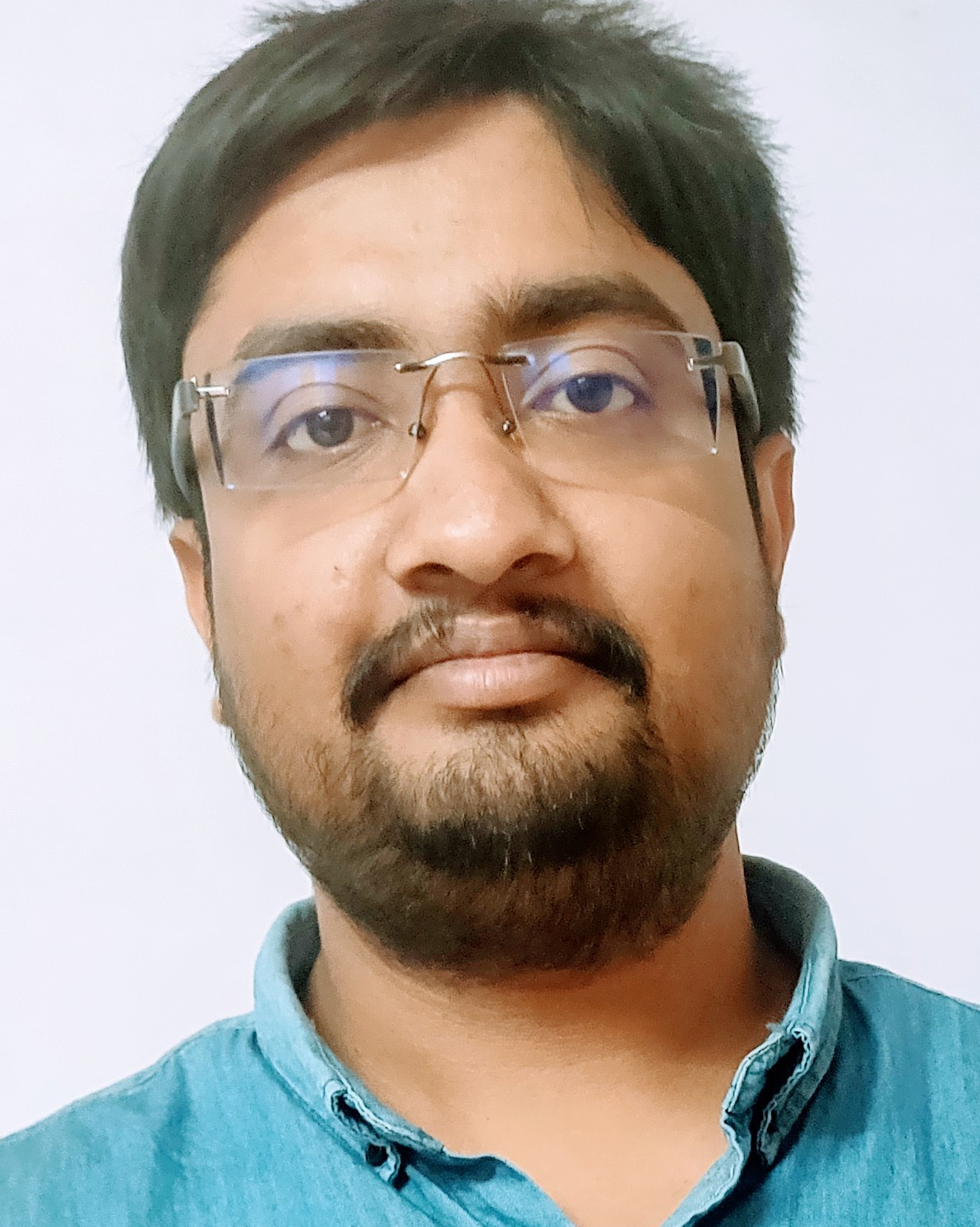}}]
    {Gururaj K} is a Masters student in the Department of Electrical Engineering at the Indian Institute of Science Bangalore. He did his B.Tech from NITK Surathkal(2013-2017). He has worked at Intel India as a software engineer. His research interests are in statistical network analysis, network representation learning, and machine learning on graphs. He is a recipient of the Cisco CSR Initiative Scholarship for Masters students.
\end{IEEEbiography}
    
\begin{IEEEbiography}
    [{\includegraphics[width=1in,height=1.25in,clip]{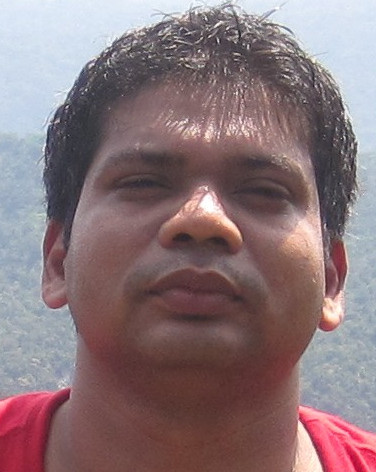}}]
    {Ambedkar Dukkipati} is an Associate Professor at the Department of Computer Science and Automation, IISc. He received his Ph.D. degree from the Department of Computer Science and Automation, Indian Institute of Science (IISc), Bangalore, India, and B.Tech from IIT Madras. He held a post-doctoral position at EURANDOM, Netherlands. 
Currently, he also heads the Statistics and Machine Learning group at the Department of Computer Science and Automation, IISc. His research interests include statistical network analysis, network representation learning, spectral graph methods, machine learning in low data regimes, sequential decision-making under uncertainty, and deep reinforcement learning.
\end{IEEEbiography}
    
\begin{IEEEbiography}
    [{\includegraphics[width=1in,height=1.25in,clip]{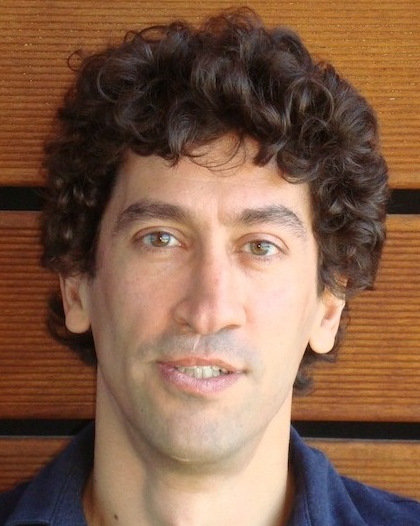}}]
    {Rui M. Castro} received a Licenciatura degree in aerospace engineering in 1998 from the Instituto Superior T\'{e}cnico, Technical University of Lisbon, Portugal, and a Ph.D. degree in electrical and computer engineering from Rice University, Houston, Texas in 2008.  Between 1998 and 2000 he was a researcher with the Communication Theory and Pattern Recognition Group, Institute of Telecommunications, Lisbon, and in 2002 he held a visiting researcher position at the Mathematics Research Center, Bell Laboratories Research.  He was a postdoctoral fellow at the University of Wisconsin in 2007-2008, and between 2008 and 2010 he held an Assistant Professor position in the department of electrical engineering at Columbia University.  He is currently an assistant professor in the Department of Mathematics and Computer Science of the Eindhoven University of Technology (TU/e) in the Netherlands.  His broad research interests include non-parametric statistics, learning theory, statistical signal and image processing, network inference, and pattern recognition.  Mr. Castro received a Rice University Graduate Fellowship in 2000 and a Graduate Student Mentor Award from the University of Wisconsin in 2008.
\end{IEEEbiography}

\end{document}